# Density of States of $Ru_3$ and $Pt_3$ Clusters Supported on Sputter-Deposited $TiO_2$

Liam Howard-Fabretto[1,2], Timothy J. Gorey[3], Guangjing Li[3], Siriluck Tesana[4], Gregory F. Metha[5], Scott L. Anderson[3], and Gunther G. Andersson[1,2]*

1 Flinders Institute for Nanoscale Science and Technology, Flinders University, Adelaide, South Australia 5042, Australia

2 Flinders Microscopy and Microanalysis, College of Science and Engineering, Flinders University, Adelaide, South Australia 5042, Australia

3 Chemistry Department, University of Utah, 315 S. 1400 E., Salt Lake City, UT 84112, United States

4 The MacDiarmid Institute for Advanced Materials and Nanotechnology, School of Physical and Chemical Sciences, University of Canterbury, Christchurch 8141, New Zealand

5 Department of Chemistry, University of Adelaide, Adelaide, South Australia 5005, Australia

*Corresponding author: Gunther G. Andersson

Email: gunther.andersson@flinders.edu.au.

Address: Physical Sciences Building (2111) GPO Box 2100, Adelaide 5001, South Australia

# Abstract

In this work, 3-atom clusters, $Ru_3$ and $Pt_3$, were deposited onto radio frequency RF-sputter deposited $TiO_2$, treated with $Ar^+$ ion sputtering. $Ru_3$ was deposited by both solution submersion and chemical vapor deposition of $Ru_3(CO)_{12}$, while $Pt_3$ was deposited under ultra-high vacuum using a laser vaporisation cluster source. The valence electronic density of states (DOS) of the deposited clusters were analysed after heat treatment using ultraviolet photoelectron spectroscopy (UPS) and metastable impact electron spectroscopy (MIES), where UPS measures the top several layers while MIES measures only the top atomic layer. XPS was used to determine the cluster surface coverages. The DOS were found to be very similar between $Ru_3$ deposited by solution submersion and chemical vapor deposition. MIES results for $Ru_3$ had contributions from titania O 2p sites due to encapsulation by a reduced titania overlayer. For $Pt_3$ clusters the UPS and MIES results provided evidence that Pt was present on the topmost layer, and encapsulation did not occur. The proposed reason for the encapsulation of $Ru_3$ but not of $Pt_3$ is the higher surface energy of Ru over Pt. It is concluded that Pt clusters deposited onto $TiO_2$ can modify the outermost layer by adding discrete energy levels on the surface, whereas the Ru clusters being encapsulated just below the surface generate a broad distribution of energy states close to the Fermi level. The outcome of this work is that $Pt_3$-cluster-modified surfaces could be used as catalysts for reactions where the $Pt_3$ energy levels are suitable for the respective reaction. The implication of the DOS found for photocatalytic water splitting are discussed.

# Introduction

Metal clusters are generally defined as groups of bound metal atoms with less than ~300 atoms [1-6]. They typically feature discrete electron energy levels similar to molecules, rather than the energy bands expected of a bulk metal counterpart including nanoparticles [7]. Clusters possess unique electronic and catalytic properties, where the addition or subtraction of a single atom to a small cluster can influence its capabilities as a catalyst [8]. Metal clusters supported on photocatalytic substrates can also act as cocatalysts by increasing the efficiency of photochemical reactions [9-11]. The catalytic and photocatalytic abilities of a cluster are related to their electronic properties) [4, 12, 13]. Thus, measurements of the valence electronic density of states (DOS) for small, supported metal clusters provide vital evidence for furthering the understanding of photocatalytic reactions at cluster surfaces and which combinations of cocatalysts and substrates are required to improve the efficiency of photocatalytic reactions.

The type of substrate used influences the catalytic properties of supported clusters [14-17]; $TiO_2$ is a photocatalytically active substrate [18] and is a common substrate choice [19-33]. Both small Ru and Pt adsorbates have been used as co-catalysts in photochemical reactions [9, 34-40]. Pt often performs well as a cocatalyst due to its high work function and capability of efficient electron migration [41, 42], but its practical applications are limited by its high cost. Ru is a cheaper cocatalyst material, but typically has less satisfactory results as a cocatalyst. Even so, some efficient $Ru/TiO_2$ photocatalysts have been produced [9], and Ru clusters have otherwise been shown to be among the most active catalysts for industry-relevant reactions such as CO and $CO_2$ hydrogenation [43-52].

One of the main ways to attach clusters onto substrates is by depositing bare clusters formed using a cluster source (CS), and deposited in high or (as here) ultra-high vacuum (UHV). A second approach is to deposit ligand-stabilised clusters such as $Ru_3(CO)_{12}$ using chemical vapor deposition (CVD) [22, 23, 53-57] or by a solution deposition (SD) approach involving submersing the substrate in a solution containing clusters [57, 58]. Each method has particular advantages. For example, SD is simple to upscale to industrially relevant scales because complex equipment is not needed and the deposition can be done on porous, high surface area supports [19, 59-61]. However, this method is less well-suited to research applications than CS or CVD [22, 23], as it is more likely to introduce contaminants to the clusters on the surface by adsorption of solvent molecules. CVD and CS, in contrast, result in surfaces with little or no external contamination but are only suitable for line-of-sight deposition and thus not suitable for modifying porous materials.

UPS and MIES can be used to measure the valence density of states (DOS) of a surface. UPS uses ultraviolet light to drive photoemission from a near-surface region limited by a combination of UV penetration and electron escape, while MIES uses thermal velocity metastable helium atoms (He*) to drive electron emission. In MIES He* does not penetrate into the surface, and rather interacts at a distance of 2 - 4 angstroms [62]. Thus MIES provides sensitivity to only the topmost surface layer of a sample, while UPS offers an information depth of approximately 2 - 3 nm [63]. When the techniques are combined, a comparison can be made between the DOS of the topmost layer and the top few nm of a surface. Additional information on MIES is provided in the Supplementary Material (page 2).

In the current work we use UPS and MIES to measure the valence DOS of $Ru_3$ and $Pt_3$ clusters deposited onto sputter-treated $TiO_2$. $Ru_3$ was deposited by both SD and CVD of $Ru_3(CO)_{12}$, and $Pt_3$ was deposited using a CS in UHV. XPS was used to determine the cluster surface coverage and chemical properties of the surfaces. The main aims are to determine the DOS of the $TiO_2$-supported $Ru_3$ and $Pt_3$ clusters and compare how the clusters interact with the $TiO_2$ and to estimate whether the DOS induced at the surface have the potential to improve the performance of photocatalysts for water splitting. Deposition of $Pt_3$ and $Ru_3$ clusters cover in the present work to cases. One of them shows the formation of discrete states and one of them metallic states. Both cases have significantly different implications for a photocatalytic process.

Low surface coverages were used to minimise cluster agglomeration, which makes DOS measurements challenging. To determine the effect of the cluster depositions on the surface DOS with suitable reliability, multiple samples were prepared for each cluster type with varied cluster loadings. To the best knowledge of the authors, there are no previous studies which have used MIES on supported $Ru_n$ or $Pt_n$ clusters. However, some UPS experiments have been performed for small $Pt_n$ clusters on various substrates [13, 64].

# Experimental

## Samples

RF-sputter-deposited $TiO_2$, shortened here to $TiO_2$, was used as the substrate for all measurements. $TiO_2$ substrates were prepared *ex situ* by RF magnetron-sputter deposition using an HHV/Edwards TF500 Sputter Coater. A 99.9% pure $TiO_2$ ceramic target was sputtered to coat P-type, boron-doped Si (100) wafers (MTI Corporation). The $TiO_2$ was treated prior to cluster deposition by heating to 723 K for 10 minutes and sputtering with $6 \times 10^{14}$ ions/cm$^2$ of $Ar^+$ at 3 keV to remove surface contamination and induce surface defects. Based on our previous measurements, the thickness of the $TiO_2$ was approximately 150 nm [65]. Further details on the RF-sputter deposition process and properties of the prepared $TiO_2$ have been provided previously [56, 65].

For each sample type, a series of multiple samples was analysed with XPS, UPS and MIES. These are herein referred to as a sample series, and each has a set of identical samples with a single changing quantity, such as the cluster surface coverage. The varying surface coverage allows the identification of the individual contributions of the substrate and clusters to the measured UPS and MIES spectra. Each metal cluster sample series featured one blank sample (with no clusters) and multiple cluster-deposited samples. The repeatability of the UPS and MIES measurements is therefore built into the measurement procedure. In addition, a sample series called Defect-$TiO_2$ was prepared with no clusters, instead featuring $TiO_2$ sputtered with 3 keV $Ar^+$ at 90° from the surface normal, with increasing doses up to a maximum of $3.6 \times 10^{15}$ $Ar^+$ ions/cm$^2$. Table 1 shows the 4 sample series which were analysed.

*Table 1: List of sample series, with series names and the physical quantities which were varied.*

| Sample Series Name | Cluster Deposition | Varied Quantity |
|---|---|---|
| Defect-$TiO_2$ | None | Substrate $Ar^+$ sputter dosage |
| SD-$Ru_3$ | SD of $Ru_3(CO)_{12}$ | Cluster coverage |
| CVD-$Ru_3$ | CVD of $Ru_3(CO)_{12}$ | Cluster coverage |
| CS-$Pt_3$ | CS of $Pt_3$ | Cluster coverage |

## Cluster Depositions

Ligated $Ru_3(CO)_{12}$ clusters were deposited using both a solution deposition process, and CVD. "Solution-deposited" (SD) $Ru_3(CO)_{12}$ clusters are referred to as "SD-$Ru_3$". To perform a solution deposition a sample was submerged into a solution of $Ru_3(CO)_{12}$ in dichloromethane for 30 minutes. Six samples of SD-$Ru_3$/$TiO_2$ and a blank $TiO_2$ sample were prepared. The solution concentration was varied from 0.002 mM to 0.2 mM to deposit a range of cluster coverages. Due to the nature of the deposition procedure, the substrate was exposed to atmosphere before and after submerging. Exposure to atmosphere was kept as short as practically possible. Samples referred to as "CVD-$Ru_3$" were prepared by depositing $Ru_3(CO)_{12}$ *in vacuo* using a CVD process which has been described previously [56]. This process involves allowing a sample of $Ru_3(CO)_{12}$ to sublime in UHV, in a geometry in which the vapor impinged and adsorbed on a $TiO_2$ substrate. Deposition times were varied from 1 minute to 90 minutes to produce a variation in cluster surface coverage. Five CVD-$Ru_3$/$TiO_2$ samples and a blank $TiO_2$ sample were prepared.

$Pt_3$ was deposited in a different UHV system (Utah), using a laser vaporisation, mass-selected CS that has been described previously [66-68]. These samples are referred to as "CS-$Pt_3$". Two CS-$Pt_3$ on $TiO_2$ samples and a blank $TiO_2$ sample were prepared. A description of the procedure is provided in the Supplementary Material (pages 2-3). All samples were heat treated as described in the next section.

**Electron Spectroscopy**

All electron spectroscopy measurements were performed using the UHV system at Flinders University. When performing measurements, UPS and MIES were performed first followed by XPS. For all spectroscopy techniques a Phoibos 100 HSA (SPECS, Germany) was used with a pass energy of 10 eV to measure the ejected electrons orthogonal to the surface. UPS, MIES, and XPS measurements were first performed on cluster samples as-deposited. The samples were then heat treated for 10 minutes before measuring again. It was found that a minimum temperature of 723 K for $Ru_3$ samples and 573 K for $Pt_3$ samples was required to achieve consistency between samples, All UPS and MIES results for the cluster deposited samples used in the present study are for the samples after heating to these temperatures. For the defect-$TiO_2$ sample series, this second heat treatment was not applied.

**XPS**

A non-monochromatic X-ray source with an Mg anode was used to produce Kα radiation with an energy of 1253.6 eV [69], at a 54.7° incidence angle to the surface. The binding energy (BE) scale was calibrated such that C 1s of adventitious carbon is 285.0 eV. Details on the XPS peak fitting and data processing procedures are given in the Supplementary Material (page 3), including details on fitting the Ru 3d/C 1s region and Pt 4f region. The XPS results were used to calculate cluster surface coverages according to the method in the Supplementary Material (page 4). The surface coverages represent the coverage of Ru or Pt atoms on the surface in units of percentage of a monolayer (% ML), with 1.0 ML defined as $2.1 \times 10^{15}$ atoms/cm$^2$ for Ru, and $1.3 \times 10^{15}$ atoms/cm$^2$ for Pt. The uncertainty in BE positions is ± 0.1 eV. Errors and uncertainties are discussed further in the Supplementary Material, pages 4-5.

**UPS and MIES**

UV light and He* were generated simultaneously by a two-stage cold cathode He discharge lamp manufactured by MFS (Clausthal-Zellerfeld, Germany). The 21.2 eV UV photons were from the He I emission line. The metastable He* was excited with 19.8 eV energy [70]. Both UV photons and He* were simultaneously directed at the surface at an angle of 54.7°, and chopped at a rate of 2000 Hz allowing for a time-of-flight separation of the UPS and MIES spectra. A bias of −10 V was applied to the sample stage for the UPS and MIES measurements to allow for recording the secondary electron cut-off as typically done for UPS and MIES measurements.

Two processes can result in electron ejection in MIES. For semiconductors or molecular materials, an Auger de-excitation (AD) process typically dominates, and usually gives spectra with discrete peak-like features. [62] MIES on metal surfaces often goes by an Auger neutralisation (AN) process and typically shows spectra with broad features. [62]

The data analysis for UPS and MIES both followed the same procedure. For BEs above 10 eV the spectra have substantial contributions from inelastically scattered secondary electrons, thus this spectral region was not included in the analysis. A linear background was subtracted from each spectrum. In each sample series, the UPS and MIES signals are analysed in terms of the contributions from the substrate and from the quantity that was varied in that series, such as the cluster surface coverage or $TiO_2$ defect density [61, 71, 72]. The spectra can be written as linear combinations of reference spectra where each reference spectrum represents one specific species (*i.e.* the clusters or substrate). The low cluster surface coverages used helps to avoid the possibility for the clusters to significantly change the DOS of the $TiO_2$ substrate (due to charge transfer, bond formation, or changes in the work function [73-75]). Details for the data analysis procedure are provided in the Supplementary Material (page 5).

For each sample series, MIE and UP spectra were analysed with component analysis. Here we have applied the algorithm singular value decomposition where a series of spectra to be analysed with SVD is considered as a matrix. In the first step the number of spectra needed to reconstruct and fit the series of spectra is determined which is given by the dimension of the matrix formed by the spectra. Analysing the dimension of the matrix results in determining the basis spectra needed to reconstruct the series of measured spectra. The basis spectra, however, have only mathematical meaning and as an example can also be negative. In the second step of the analysis a linear combination of the basis spectra is applied to determine physically meaningful reference spectra. This operation is applied with three conditions: i) the reference spectra must be non-negative, ii) the weighting factors for fitting the measured spectra as linear combination of the reference spectra must be non-negative and iii) the sum of the weighting factors to fit a measured spectrum must be unity. Details of the SVD analysis have been described previously. [76, 77]

## Results

### XPS

For all samples, the Ru 3d/C 1s region, Ti 2p region, and O 1s region were measured with XPS, and for $Pt_3$ samples the Pt 4f region was also scanned. *Figure 1* shows overlayed XP spectra for each sample series. For Defect-$TiO_2$, the Ti 2p region is shown to highlight the increasing size of the $Ti^{3+}$ and $Ti^{2+}$ peaks with increasing $Ar^+$ sputter dosage as can be seen in the 457 – 458 eV region with details described further below. $Ti^{4+}$ is related to stoichiometric titania, while $Ti^{3+}$ and $Ti^{2+}$ relate to defects in the $TiO_2$ structure. For $Ru_3$ and $Pt_3$ cluster samples, the Ru 3d/C 1s region and Pt 4f region are shown, respectively. It was confirmed by the areas of the Ru 3d and Pt 4f peaks that there was a range of cluster surface coverages between the samples for each series. Examples of the peak fitting used in the Ti 2p, Ru 3d, and Pt 4f regions are shown in Figure 2, where the spectra were fitted to determine the cluster surface coverages (Table 2) and core BEs (Table 3) for each sample. The fitted spectra for blank samples (without deposited clusters) are provided in the Supplementary Material (Figure S1).

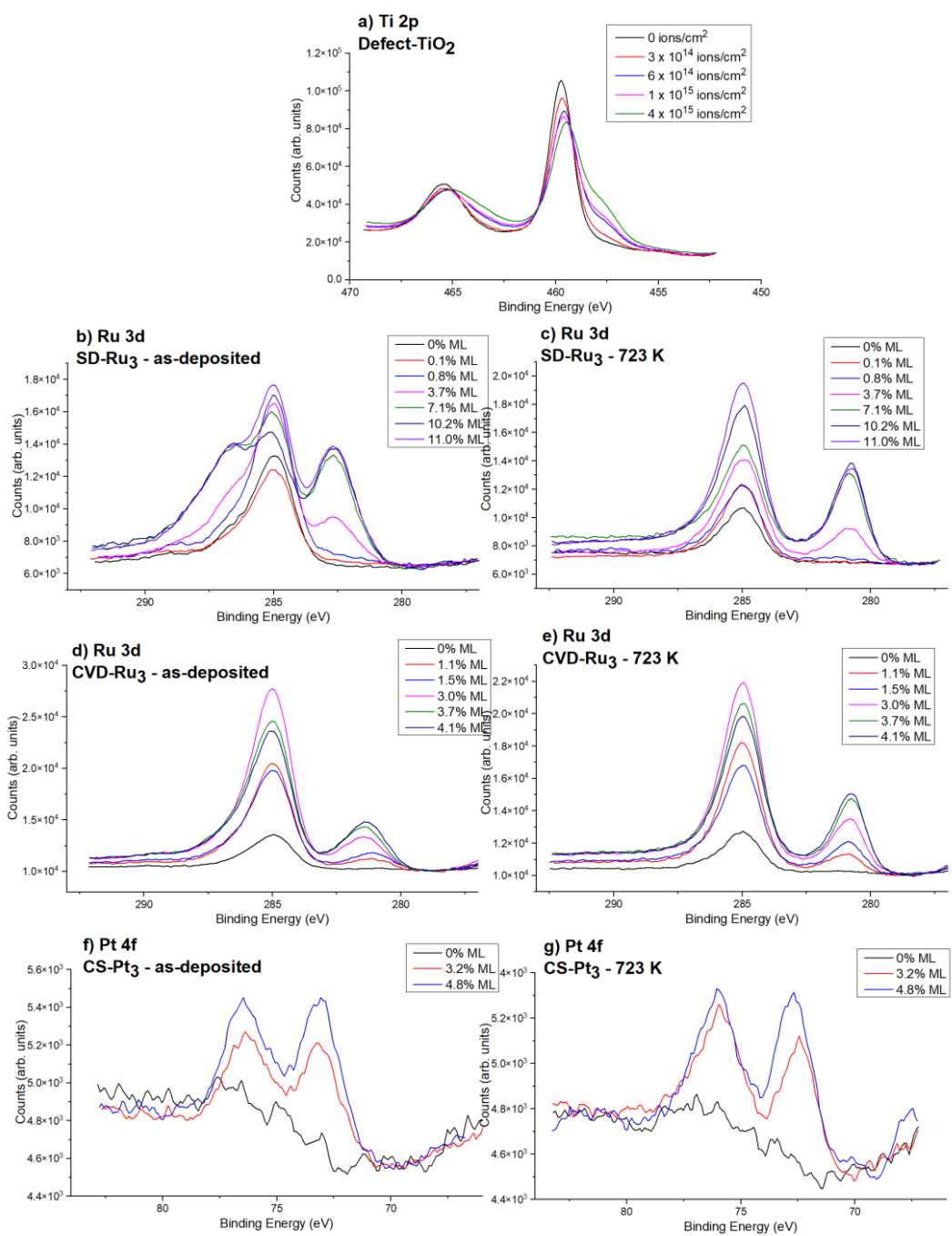

*Figure 1: XPS results with spectra from the same sample series overlayed. a) Ti 2p region for Defect-TiO$_2$. b) Ru 3d region for SD-Ru$_3$ as-deposited. c) Same samples as (b) after 723 K heat-treatment. d) Ru 3d region for CVD-Ru$_3$ as-deposited. e) Same samples as (d) after 723 K heat-treatment. f) Pt 4f region for CS-Pt$_3$ at room temperature. g) Same samples as (f) after 573 K heat-treatment.*

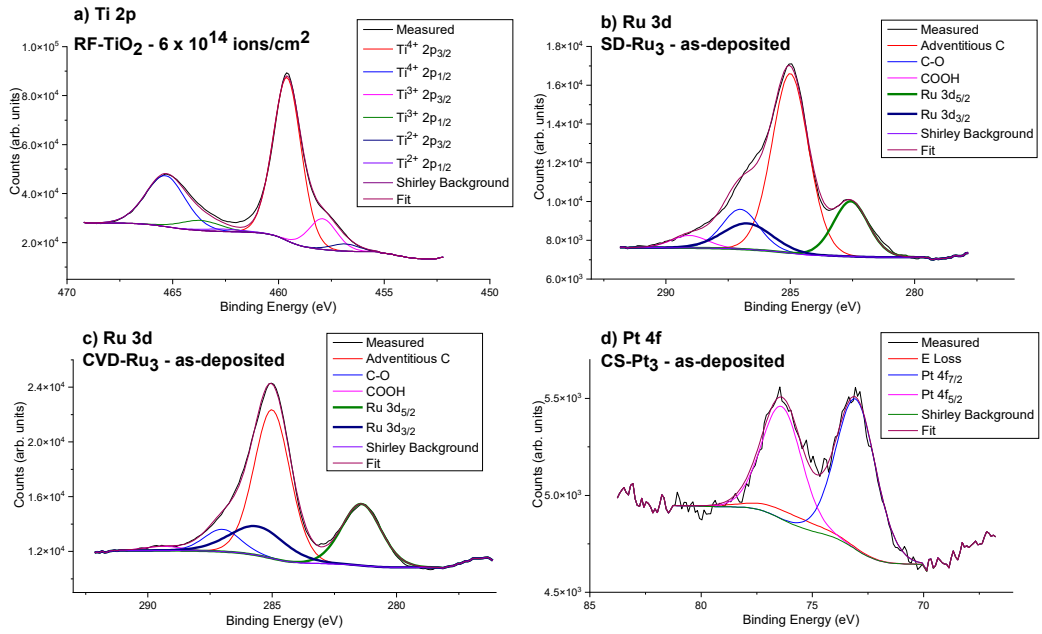

*Figure 2: XPS peak fitting examples. a) Ti 2p peak fitting for $TiO_2$ after heating and sputtering. b) Ru 3d/C 1s peak fitting for SD-$Ru_3$ as-deposited (0.1 mM sample). c) Ru 3d/C 1s peak fitting for CVD-$Ru_3$ as-deposited (30 min deposition). d) Pt 4f region fitting for CS-$Pt_3$ sample (4.8% ML) at room temperature.*

*Table 2: Cluster surface coverages, determined from XPS measurements of as-deposited cluster samples at room temperature. The absolute error in the surface coverage is ~100%, while the relative uncertainty when comparing samples is only ± 4% (see Supplementary Material pages 4-5 for details).*

| Sample Series | Sample | Surface Coverage (% ML) |
|---|---|---|
| **SD-Ru$_3$** | SD-Blank | 0.0 |
| | SD-Ru$_3$-1 | 0.1 |
| | SD-Ru$_3$-2 | 0.8 |
| | SD-Ru$_3$-3 | 3.7 |
| | SD-Ru$_3$-4 | 7.1 |
| | SD-Ru$_3$-5 | 10.2 |
| | SD-Ru$_3$-6 | 11.0 |
| **CVD-Ru$_3$** | CVD-Blank | 0.0 |
| | CVD-Ru$_3$-1 | 1.1 |
| | CVD-Ru$_3$-2 | 1.5 |
| | CVD-Ru$_3$-3 | 3.0 |
| | CVD-Ru$_3$-4 | 3.7 |
| | CVD-Ru$_3$-5 | 4.1 |
| **CS-Pt$_3$** | CS-Blank | 0.0 |
| | CS-Pt-1 | 3.2 |
| | CS-Pt-2 | 4.8 |

Figure 2a shows the Ti 2p region for TiO$_2$ after sputtering with 6 x 10$^{14}$ Ar$^+$ ions/cm$^2$. The titania regions have been fitted with three doublets relating to Ti$^{4+}$, Ti$^{3+}$, and Ti$^{2+}$. Ti$^{4+}$ was located at 459.6 eV ± 0.1 eV, and Ti$^{3+}$ and Ti$^{2+}$ were shifted by -1.7 eV ± 0.1 eV and -2.7 eV ± 0.1 eV, respectively. It should be noted, however, that the background in the Ti 2p region is not structureless, adding uncertainty to the fitting process. For example, an apparent Ti$^{2+}$ doublet is also present for unsputtered TiO$_2$ (Figure S1a) even though this sample was air exposed prior to analysis, which would tend to oxidize reduced sites in the surface. Thus, it is likely that the small Ti$^{2+}$ doublet feature fit for the sputtered samples is partially an artifact of the varying background. Table S1 shows the relative atomic concentration of Ti$^{4+}$, (Ti$^{3+}$ + Ti$^{2+}$), and Ru or Pt for each sample. Within the XPS information depth, the relative concentration of Ti$^{3+}$ and Ti$^{2+}$, related to defected titania, was unrelated to the presence of clusters and was relatively consistent between the samples.

Figure 2b and c show the Ru 3d region for the SD-Ru$_3$ and CVD-Ru$_3$ samples. Note that in addition to the Ru 3d doublet, this region also has a substantial C 1s peak due to adventitious carbon, overlapping the higher energy Ru 3d$_{3/2}$ peak. Therefore, the Ru spectral fits are based on the Ru$_{5/2}$ peak, for which the carbon contribution is negligible, with the Ru $3d_{3/2}$ peak constrained to be in a 2:3 intensity ratio. The $Ru_{5/2}$ peak was in the 281-283 eV range for the unheated samples, shifting to lower BEs when the sample is heated, due to the loss of ligands. Figure 2d shows the Pt 4f region for CS-$Pt_3$. An energy loss (E loss) peak was present at ~75 eV on both the blank (Figure S1c) and $Pt_3$-loaded samples. In addition to the E loss peak there was a Pt $4f_{7/2}$ and $4f_{5/2}$ doublet. Ru 3d peaks were fit with an asymmetrical peak shape, all other peaks were fit with symmetrical peaks.

*Table 3: XPS core electron BEs for Ru $3d_{5/2}$ and Pt $4f_{7/2}$. XPS was measured as-deposited and after heat-treatment. The BE shift is the difference in BE before and after the heating procedure. The BE uncertainty is ± 0.1 eV.*

| Sample Series | Peak | Initial BE (eV) | Heat-Treated BE (eV) | BE Shift (eV) |
|---|---|---|---|---|
| SD-$Ru_3$ | Ru $3d_{5/2}$ | 282.6 | 280.7 | -1.9 |
| CVD-$Ru_3$ | Ru $3d_{5/2}$ | 281.4 | 280.8 | -0.6 |
| CS-$Pt_3$ | Pt $4f_{7/2}$ | 73.0 | 72.5 | -0.5 |

The BEs of the main metal cluster peaks for each sample series, determined by averaging the values for the different coverage samples in each series, are shown in Table 3. The SD-$Ru_3$ clusters have a higher BE than CVD-$Ru_3$ deposited before heat treatment, but after heat treatment both types of $Ru_3$ cluster samples shifted lower in BE to 280.7 eV ± 0.1 eV and 280.8 eV ± 0.1 eV, respectively. The reported BE after heating is higher than bulk Ru, which has a BE of 279.7 eV ± 0.2 eV.[57] The heating-induced BE shifting of -1.9 ± 0.1 eV for SD-$Ru_3$, and -0.6 eV ± 0.1 eV for CVD-$Ru_3$, are predominantly due to the removal of CO ligands from the clusters [22, 23]. The binding energy shifts results are comparable to those reported for SD-$Ru_3$ and CVD-$Ru_3$ on $TiO_2$ in our previous study [57]. This suggests the properties are comparable between the samples in the present study and our previous study. The XPS BE results are also aligned with a previous study by Zhao *et al.* [22] on $Ru_3(CO)_{12}$ deposited by CVD onto $TiO_2$(110).

Table 3 shows that the Pt $4f_{7/2}$ peak for CS-$Pt_3$ was present at 73.0 eV ± 0.1 eV at room temperature and shifted -0.5 eV to 72.5 eV ± 0.1 eV after heating to 573 K. Because the $Pt_3$ clusters were deposited *ex situ*, it is likely the -0.5 eV ± 0.1 eV peak shift after heating was related to the removal of contamination from the surface of the clusters. These values are outside the range of typically reported bulk Pt BEs, i.e. 70.9 to 71.3 eV [57]. The increase in BE for cluster over bulk Pt may be due to a final state effect related to the small cluster size, as there is a trend for increasing core level BE with decreasing cluster size, due to less final state stabilisation by screening and charge delocalisation [29, 78, 79]. The Pt $4f_{7/2}$ BE for small Pt clusters being greater than that of bulk Pt aligns with previous studies [13, 29, 79-83]. However, the BE is slightly greater than 71.9 eV reported in previous studies of the same-sized $Pt_3/TiO_2$(110), [80] and 71.8 eV for $Pt_2/SiO_2$ [81]. The surface coverages used in those studies were 5-10%, which are similar to this study, so differences in cluster-cluster interactions do not account for the BE differences. This higher BE is also not related to fully oxidised Pt clusters because the BE is lower than that of bulk PtO, which is reported to be ~74.0 eV [84]. Thus, the higher Pt 4f BE for CS-Pt$_3$ in this study may be related to a unique cluster-surface interaction for Pt$_3$ on the RF-sputter-deposited TiO$_2$.

**UPS and MIES**

For each UPS/MIES series, a range of spectra were measured both before and after heating (723 K for Ru$_3$ samples and 573 K for Pt$_3$ samples) but only the latter are reported. The measurements before heating are not included in the analysis because the CO ligands are still present for the SVD-Ru$_3$ samples and also some impurities due to the surfaces being exposed to atmosphere making it too difficult to separate the contributions in the electron spectra related to the impurities and the CO from those related to the substrate and the clusters. For each series, the measured UPS and MIES spectra for each sample are shown as well as the calculated reference spectra as determined through the SVD analysis described in the Experimental section. Weighting factors vs Ti defect ratio or cluster surface coverage are shown in the Supplementary Material (Figure S2). Examples of the fitted spectra are shown in the supplementary section for the UP spectra of CVD-Ru3 in Figure S3 and Figure S4. A 99.9% pure metallic Ru reference sample was also measured with UPS and MIES (see Supplementary Material, Figure S3). UPS and MIES data for SD-Ru$_3$, CVD-Ru$_3$, and CS-Pt$_3$ are shown in Figure 3, Figure 4, and Figure 5 respectively. When compared to the TiO$_2$ blank samples, the cluster depositions made visible differences to the measured UPS and MIES spectra, which are attributed to the UPS and MIES signal from the deposited metal clusters on the surfaces. The Defect-TiO$_2$ sample series was prepared to determine how the Ar$^+$-induced defects affect the UPS and MIES spectra of the TiO$_2$ surface, with results shown in Figure 6.

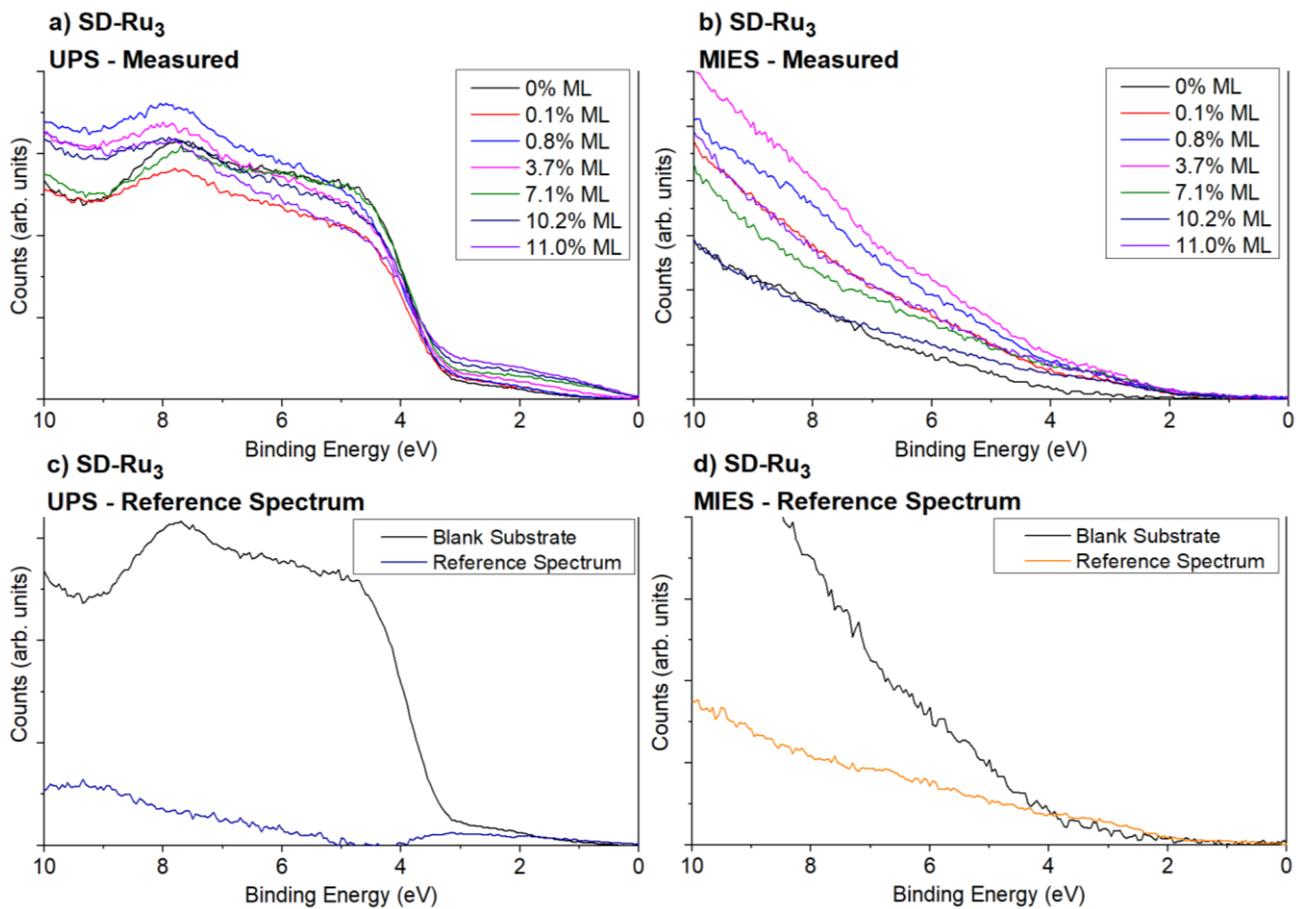

*Figure 3: UPS/MIES results for the SD-Ru$_3$ sample series after heating to 723 K. a) Measured UPS spectra. b) Measured MIES spectra. c) SD-Ru$_3$ UPS reference spectrum. d) SD-Ru$_3$ MIES reference spectra.*

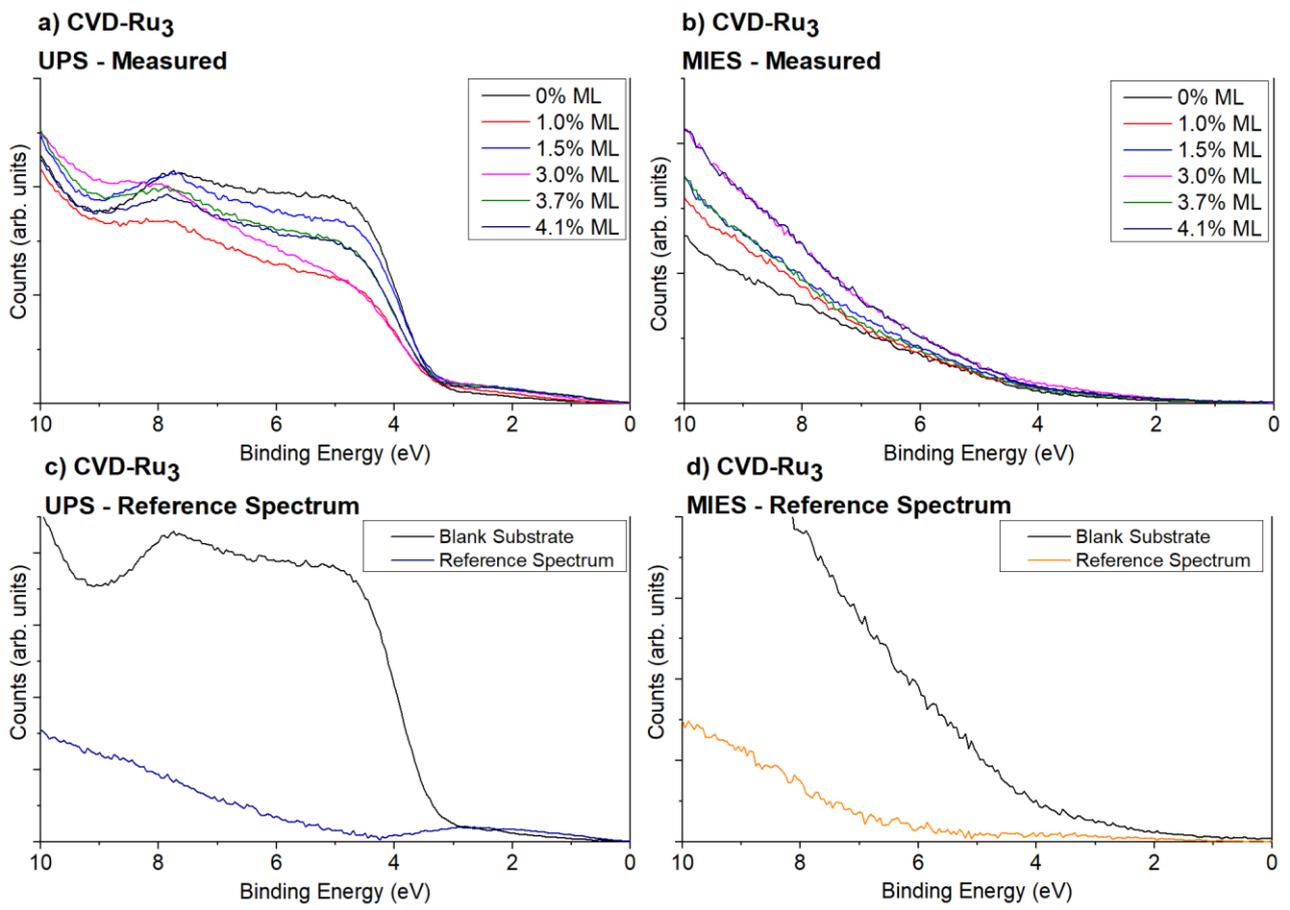

*Figure 4: UPS/MIES results for the CVD-Ru$_3$ sample series after heating to 723 K. a) Measured UPS spectra. b) Measured MIES spectra. c) CVD-Ru$_3$ UPS reference spectra. d) CVD-Ru$_3$ MIES reference spectra.*

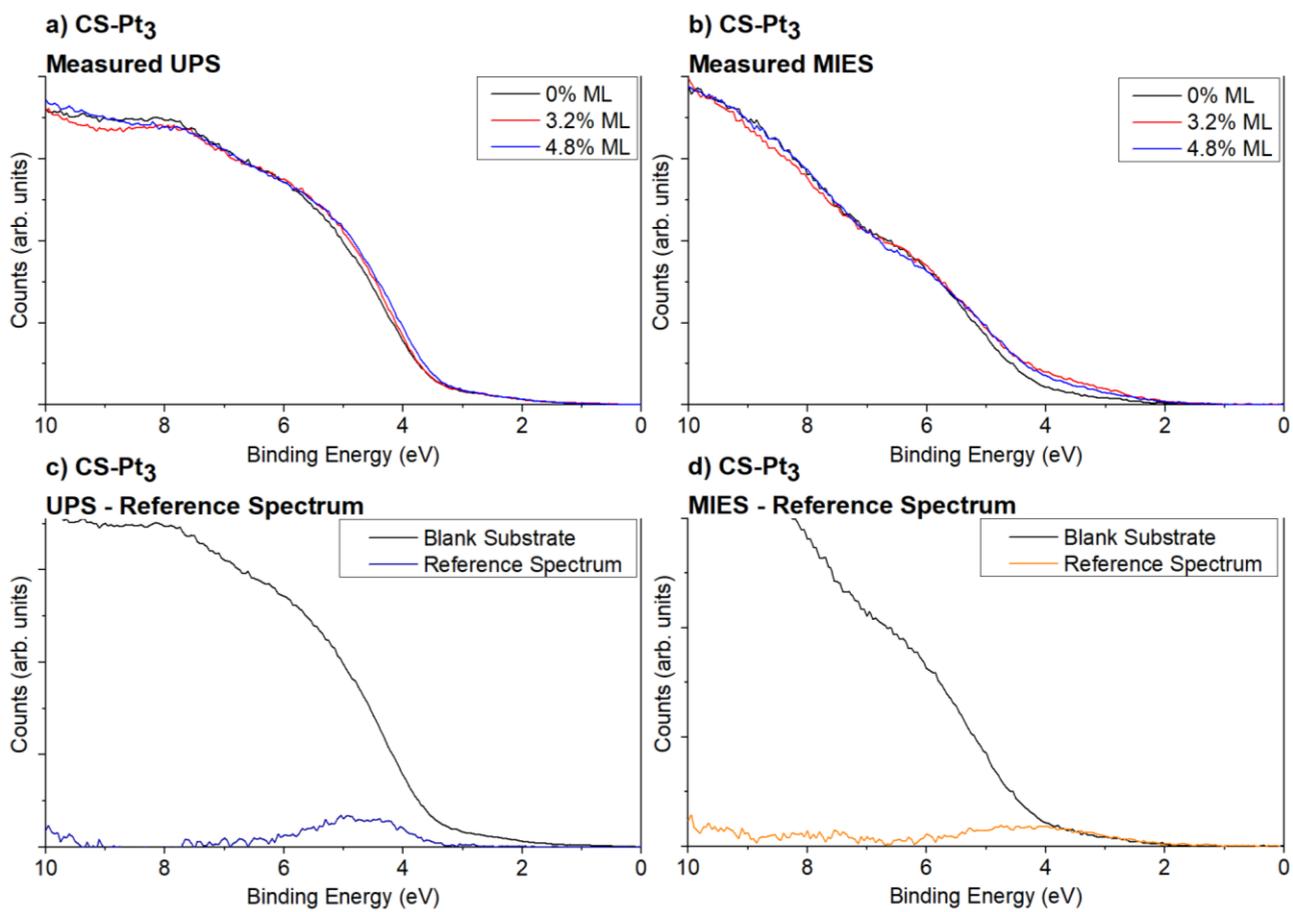

*Figure 5: UPS/MIES results for the CS-Pt$_3$ sample series after heating to 573 K. a) Measured UPS spectra. b) Measured MIES spectra. c) CS-Pt$_3$ UPS reference spectra. d) CS-Pt$_3$ MIES reference spectra.*

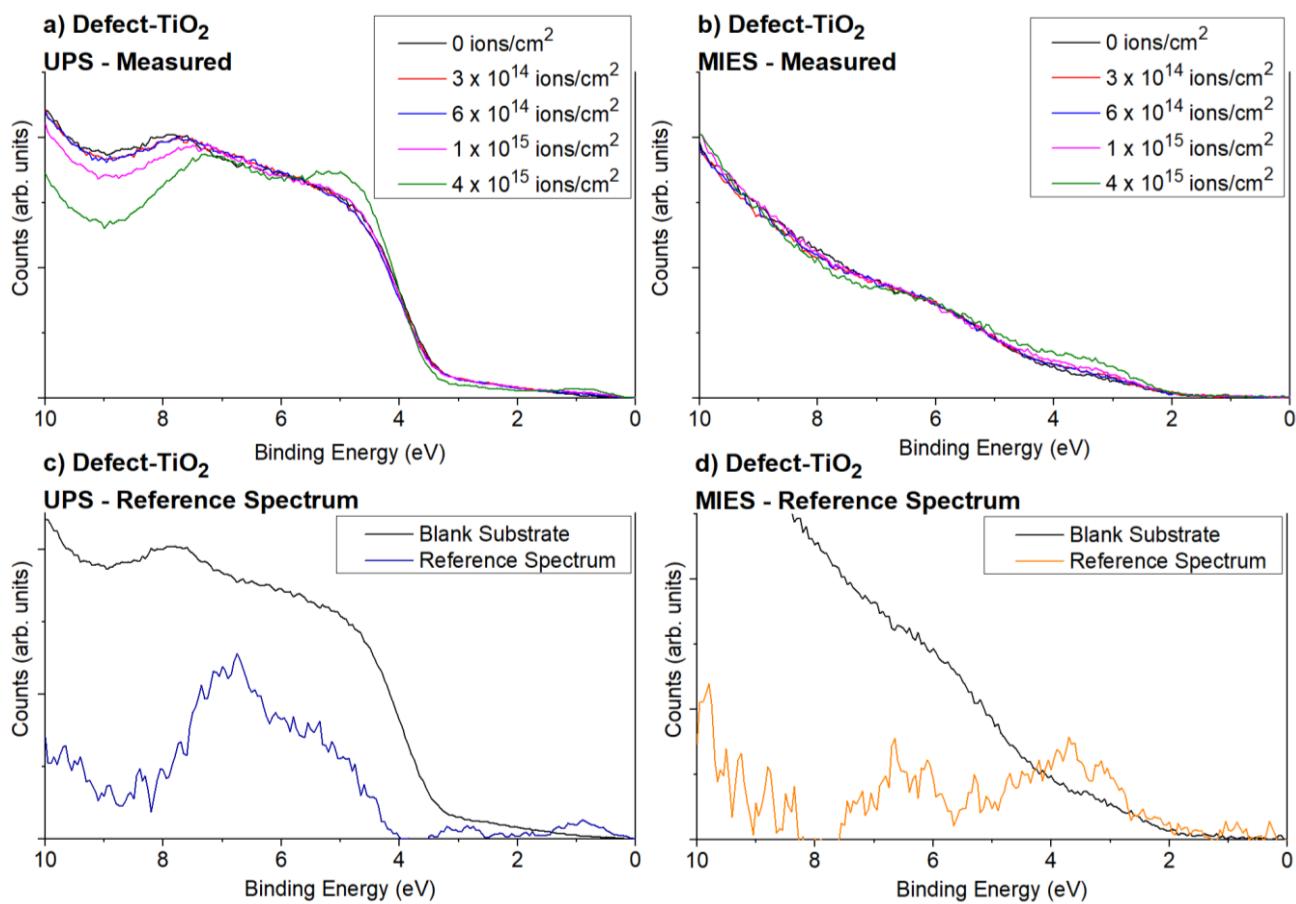

*Figure 6: UPS/MIES results for Defect-TiO$_2$. a) Measured UPS spectra. b) Measured MIES spectra. c) Defect-TiO$_2$ UPS reference spectra. d) Defect-TiO$_2$ MIES reference spectra.*

The main feature for the SD-Ru$_3$ UPS reference spectrum (Figure 3c) is an asymmetrical energy band from 0 to 4.0 eV, with a peak located at 3.1 eV. For MIES (Figure 3d), the SD-Ru$_3$ reference spectrum has a feature at approximately 3.3 eV, but the peak is weak and somewhat obscured by the secondary electron background. The main feature for the CVD-Ru$_3$ UPS reference spectrum (Figure 4c) is an asymmetrical energy band from 0 to 4.1 eV, with a peak at 2.7 eV. For MIES (Figure 4d), the CVD-Ru$_3$ reference spectrum has a feature at approximately 3.6 eV. By comparison to other Ru materials, it is likely that the main features of these spectra represent Ru 4d states [85-87] which can also be seen in Figure S3, for bulk Ru metal. The UPS and MIES reference spectra are very similar between SD-Ru$_3$ and CVD-Ru$_3$. The main feature of the CS-Pt$_3$ UPS reference spectrum in Figure 5c is at 4.7 eV, and is likely due to 5d states [13, 88, 89]. There do not appear to be any other peaks in the spectrum, and there is no signal below or above this peak (besides signal at higher BEs, related to the ejection of secondary electrons). In the MIES reference spectrum (Figure 5d) there is a single peak at 4.3 eV. The peak is similar to that seen in UPS but shifted by -0.4 eV and broadened. UPS of Pt particles grown on alumina was reported by Altman and Gorte as a function of Pt coverage.[90] For large particles at high coverage, a bulk-like spectrum was observed, with high intensity at the Fermi level. With decreasing coverages (smaller particles), the Pt-associated intensity dropped and shifted to higher BE, such that for ~2.5 nm particles, the onset was ~1 eV below the Fermi level. Thus, the observation of an onset more than 2 eV below the Fermi level for CS-Pt$_3$ suggests that the Pt$_3$ clusters did not sinter extensively to product large particles. MIES of bulk Pt would most likely have a broad and featureless spectrum near 0 eV binding energy.[62]

The shape of the Defect-TiO$_2$ spectra for UPS and MIES in Figure 6a-b changed as the Ar$^+$ ion sputter dose was increased, with the most dramatic change being for the 3.6 x 10$^{15}$ ions/cm$^2$ spectrum. From SRIM calculation it can be estimated that there is preferential sputtering of O leading to a total deficit of about 1.1 x 10$^{15}$ O atoms/cm$^2$ which is about the number of units of TiO$_2$ per cm$^2$. The main features are the same for all UPS spectra (Figure 6a), and before sputtering are present at 0.8 eV, 5.0 eV and 8.0 eV. The 0.8 eV feature in the band gap is most prominent after higher sputter doses; this peak is related to Ti$^{3+}$ defects [18, 29]. The other peaks are related mostly to O 2p levels [91, 92] and are part of the titania valence band, which here spans from ~3 to 9 eV. After sputtering at 3.6 x 10$^{15}$ ions/cm$^2$ the 8.0 eV BE peak shifted lower by 0.8 eV. Interestingly, the spectral intensity increased at ~7 eV with sputter dosage, but then decreased somewhat for the largest sputter dose; this may suggest that there is a more significant change in O 2p structure after reaching a large enough sputter dosage. The MIES spectra (Figure 6b) has broader features at 3.5 eV and 6.6 eV, which appear to be shifted versions of the two higher energy UPS features. Due to the surface sensitivity of MIES, these are likely modified O 2p states due to surface defects. No Ti$^{3+}$ feature at ~0.8 eV is seen in MIES, most likely because Auger neutralisation is favoured at the associated orbitals [62]. The reference spectra (Figure 6c-d) represent the components of the defected titania toward the measured spectra. The UPS reference spectrum matches previous experiments which have been performed on amorphous Ti$^{3+}$-doped TiO$_2$ nanoparticles [93].

## Discussion

**Reference Spectra**

The UPS and MIES reference spectra (treatments described in the Experimental section) as determined from the SVD analysis for Defect-TiO$_2$, SD-Ru$_3$, CVD-Ru$_3$, and CS-Pt$_3$ are overlaid in Figure 7 for comparison. These are the spectra extracted by fitting each sample series, and represent the spectra of the species being varied in each series, i.e., the spectra of the TiO$_2$ defects or the deposited clusters, depending on the series. The reference spectrum for each sample series as determined from the SVD analysis was multiplied by a linear scaling factor due to differing count rates, which means the absolute intensities of the spectra cannot be directly compared. In addition, due to the cross sections being unknown we cannot determine species concentrations from the UPS and MIES results.

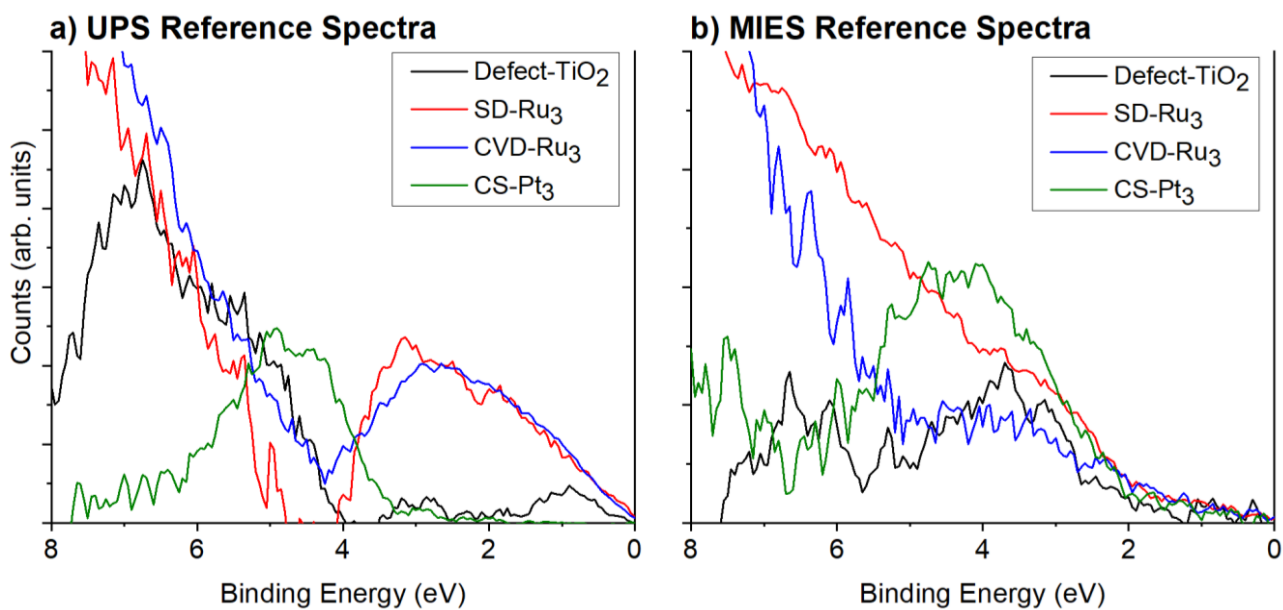

Figure 7: Overlayed UPS (a) and MIES (b) reference spectra for Defect-TiO$_2$, SD-Ru$_3$, CVD-Ru$_3$, and CS-Pt$_3$.

For SD-Ru$_3$, the UPS (Figure 7a) and MIES (Figure 7b) reference spectra both have single peak and shoulder like structures with similar locations near 3eV, where the UPS peak is more easily identifiable. This is also the case for CVD-Ru$_3$. The main differences are that the MIES spectra in each case have a larger contribution from the secondary electron background (especially SD-Ru$_3$), and a far lower intensity between 0-2 eV. For CS-Pt$_3$ the main feature around 4-5 eV is very similar between UPS and MIES, and the main difference is an earlier onset for MIES. Notably, for Defect-TiO$_2$ the spectrum is very different between UPS and MIES. This difference suggests that He* is de-exciting via Auger neutralisation (AN) for Defect-TiO$_2$ which can broaden or alter the spectrum.

It needs to be considered why in Figure 7 the peak in the MIE spectrum for the CS-Pt$_3$ clusters is shifted by ~0.8 eV to lower binding energy compared to the UP spectrum. The reason is most likely an image charge effect for the He* interaction with the surface. [62] It also can be seen that the MIE spectrum of SD-Ru$_3$ has little structure compared to the UP spectrum, which we attribute to the Auger de-excitation (AD) mechanism dominating for He* interacting with a metallic species like the agglomerated Ru clusters. This effect can also be seen in Figure S3. The dominance of AN vs AD mechanisms for He* interacting with semiconducting vs metallic surfaces is described in the Experimental section.

Comparing UPS spectra in more detail, SD-Ru$_3$ and CVD-Ru$_3$ (Figure 7a) have the same features with similar BE dependence. The MIES spectra for these samples (Figure 7b) have similar onsets to each other, and both show evidence of a feature between 2 and 5 eV, which, however, is much better resolved in the CVD-Ru$_3$ spectrum. The XPS results (Table 3) showed that after heat treatment to 723 K, the Ru 3d BE is the same (± 0.1 eV) for these two sample types. Heating to 723 K involves complete or near-complete ligand removal from the CVD-Ru$_3$ deposited clusters and makes them similar to the SD-Ru3 clusters [56], which suggests that the ligand-removal step has a greater effect than the deposition method on the electronic properties of the clusters.

The MIES reference spectra for SD-Ru$_3$ and CVD-Ru$_3$ spectra, while significantly different, both show features in the same DOS region as the Defect-TiO$_2$ MIES spectrum (Figure 7b). All three spectra feature a peak in the 3.0 to 3.5 eV region, which for the Defect-TiO$_2$ reference spectrum was assigned to defect-modified O 2p sites. This suggests that both the SD-Ru$_3$ and CVD-Ru$_3$ MIES reference spectra contribute to a cluster-induced increase in surface-layer titania defects by modifying the DOS of TiO$_2$. Previous studies have also shown increases in titania defect sites due to the encapsulation of metals by titania [21, 24, 94-101], which supports this interpretation. Notably, it is possible to consider that Ru clusters are contributing to the MIES reference spectra in addition to the O 2p sites, because there are also peaks in a similar location in the SD-Ru$_3$ and CVD-Ru$_3$ UPS spectra (Figure 7a). However, in our previous study it was shown by temperature-dependent low energy ion scattering (TD-LEIS) that there is no Ru available on the surface layer after heating to 660 K ± 120 K, where the clusters were encapsulated by an average of 0.35 nm ± 0.08 nm, which is approximately 1.7 ML of titania [56, 57]. Ru was also present deeper in the substrate as confirmed by STEM, however these Ru particles could not be detected by electron spectroscopy due to their depth [57]. This suggests that while Ru contributes to the MIES spectra, the features also might have contributions from modified O 2p sites. With the available data it is difficult to separate the specific Ru features from the modified O 2p sites. Conversely, for the UPS reference spectra (Figure 7a) the Defect-TiO$_2$ spectrum is not like that of SD-Ru$_3$ or CVD-Ru$_3$, and thus the UPS reference spectra determined do not represent titania O 2p sites. This is most likely due to the greater probing depth of UPS, approximately 2-3 nm [63], which is deep enough to probe Ru clusters encapsulated by 0.35 nm ± 0.08 nm of substrate material [57].

The shape of the CS-Pt$_3$ reference spectra are unique compared to the other UPS/MIES spectra (Figure 7). Furthermore, the UPS and MIES spectra are similar to each other; the only difference being some broadening towards the Fermi level in MIES, with an accompanying −0.4 eV peak shift. The CS-Pt$_3$ MIES low-BE onset is at a similar location to the Defect-TiO$_2$ MIES onset. Thus, the peak broadening towards the Fermi level for the CS-Pt$_3$ MIES reference spectrum may be due to some contribution of titania surface-layer defects to the MIES spectrum. The presence of Pt in the MIES spectrum provides evidence that the Pt$_3$ clusters remain present on the topmost layer, which is in contrast to SD-Ru$_3$ and CVD-Ru$_3$. The presence of Pt on the topmost layer is most likely a benefit for photocatalysis, because photocatalytic reactions typically occur at active sites on the topmost surface layer [102]. It is worth noting that Pt was found to encapsulate into TiO$_2$(110) by heating to 673 K [90] and 773 K[103], respectively, i.e. at temperatures of 100 to 200 K higher compared to those used here.

**Cluster-Surface Interaction**

For both SD-Ru$_3$ and CVD-Ru$_3$ the UPS DOS onset is at the Fermi level (0 eV in Figure 7a), which is also the case for the metallic Ru reference UPS (Figure S3). This could suggest that the clusters have metallic states [64, 104-107], which would be surprising for such small clusters, which typically have molecule-like discrete electronic levels, rather than metallic properties [7]. Our previous studies [57] showed that Ru$_3$ clusters were partially oxidised and fully encapsulated after heating to 800 K on TiO$_2$ with some of the clusters agglomerated and forming small Ru nanoparticles with diameters up to 1.5 nm. This points towards a strong metal-support interaction (SMSI) occurring. Also, small Ru nanoparticles have metallic character and would explain the UPS DOS onset to be found at 0 eV. The SMSI can occur for adsorbates such as transition metal clusters when supported on reducible metal oxide surfaces including TiO$_2$, and can involve cluster oxidisation or charge transfer [21, 108, 109], as well as encapsulation by the substrate, typically after high temperature reduction [20, 21, 24, 94-99, 110-114]. It is possible that metallisation of the Ru$_3$ clusters on TiO$_2$ results from the SMSI. TiO$_2$ defects also result in appearance of a peak just below the Fermi level (Fig. 7a and [18, 29]), however it is unlikely the DOS observed near the Fermi level in UPS for the cluster samples is due to the presence of TiO$_2$ defects because XPS showed no correlation between the presence of clusters and the amounts of Ti$^{2+}$ and Ti$^{3+}$ (see Table S1). For CS-Pt$_3$ the DOS onset was at 3.3 eV for UPS and 2 eV for MIES. Considering the onset of the Pt$_3$ DOS is not close in energy to the Fermi level and does not exhibit the modified step function characteristic of a metallic Fermi level, i.e., the supported Pt$_3$ clusters retained molecule-like electronic properties, suggesting that aggregation to larger particles with metallic properties did not occur [64, 105-107]. Such a conclusion is aligned with previously reported UPS results for Pt$_n$ clusters [13, 64], and is in contrast to bulk forms of Pt such as Pt(111) where the UPS onset is at the Fermi level [89]. The similarity between the UPS and MIES spectra suggests that He* de-excitation in MIES was predominantly by the AD mechanism, providing further evidence that the clusters have non-metallic properties [62]. This notion is also supported by previous studies showing Pt$_n$ clusters have a DOS onset which shifts closer to the Fermi level as cluster size increases, including a UPS study of Pt$_n$ (n = 1-15) on Ag(111) [115], and a DFT study of Pt$_n$ (n = 1-4) on TiO$_2$(110) [116]. The non-metallic properties of the Pt clusters are further supported by the Pt 4f XPS results (Figure 2d) which feature symmetrical Pt doublet peak shapes with large final state shifts to high BE, contrasting bulk Pt where the 4f doublet typically has asymmetrical peaks [80]. A previous study by Isomura *et al.* [80] has similarly shown symmetrical Pt 4f peak shapes for small, size-selected Pt$_n$ clusters.

Several previous studies have been performed analysing the DOS of small, size-selected Pt$_n$ clusters using UPS [13, 64] on other substrates, but to the best knowledge of the authors this is the first study using TiO$_2$. Comparisons between this study and previous studies are therefore important to determine how the substrate effects the DOS. The UPS and MIES reference spectra for Pt$_3$ are comparable to the UPS reference spectrum determined in a study by Eberhardt *et al.* [64] for CS-deposited, size-selected $Pt_3/SiO_2$; this study found two peaks were present at ~4.9 eV and ~8.5 eV. The 4.9 eV peak appears to be the same feature as the 4.7 eV peak in the UPS for $CS-Pt_3$ (Figure 7a). The second, higher BE feature was not present in this study, which may be due to differences in the cluster-surface interaction, or an effect of variations in secondary electron contributions to the reference spectra.

As discussed, $SD-Ru_3$ and $CVD-Ru_3$ deposited onto $TiO_2$ were encapsulated by a layer of reduced titania after heating to 723 K treatment. [57] Conversely, $CS-Pt_3$ was not encapsulated by $TiO_2$ based on the MIES – being sensitive to only the outermost layer of the sample – results identifying DOS which can be related to $Pt_3$ and also based on the observation that the MIES and UPS results are quite similar, even after heating to 573 K. This suggests that the SMSI is at least stronger for $Ru_3$ clusters compared to the $Pt_3$ clusters. The higher heating temperature is not a likely reason for the differences in the encapsulation process between the samples because encapsulation into $TiO_2$ was also found for $Ru_3$ at 573 K. As noted above, encapsulation of Pt into $TiO_2$ was found previously at 673 K [90] and 773 K [103].

Figure 8 shows a diagram summarising our hypothesis regarding encapsulation for $Ru_3$ and $Pt_3$ on sputtered $TiO_2$ after heating with all core findings summarised in Table 4. Previous studies have shown Pt adsorbates being encapsulated by $TiO_2$(110) due to heating, which is in opposition to these result, however these were for larger Pt nanoparticles (not small clusters) and/or much higher surface coverages than those used in this study [95, 99, 117]. In a study by Wu *et al.*, it was suggested that supported Pt clusters on $TiO_2$(110) were not encapsulated during temperature programmed deposition (TPD), while larger supported nanoparticles were encapsulated due to increased SMSI [117]. This study put forward the idea that the encapsulation of Pt adsorbates by titania is promoted by larger Pt particle sizes, which is in alignment with the present study showing that $Pt_3$ clusters were not encapsulated. Similarly, ultrafine Pt nanoparticles (~1 nm) on $TiO_2$ powder have been shown to resist encapsulation after heating to 723 K under UHV [118].

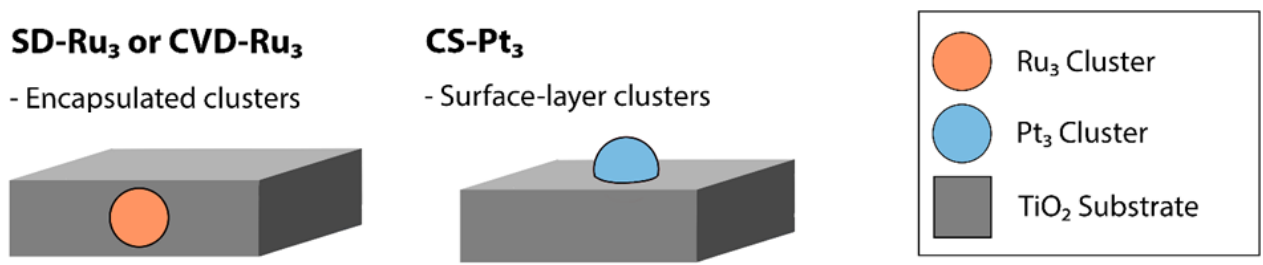

*Figure 8: Summary of the results regarding encapsulation for $SD-Ru_3$, $CVD-Ru_3$ and $CS-Pt_3$ clusters on $TiO_2$. These images represent the surfaces after heat treatment. Only one image is shown for $Ru_3$ clusters because the same results were found for both deposition methods.*

The mechanism for the encapsulation of $Ru_3$ (but not $Pt_3$) on $TiO_2$ substrates is most likely related to the minimisation of surface energy in each of the systems [21, 94, 96, 119]. The surface free energy for Ru (~2.7 J.m$^{-2}$) is higher than that of Pt (~2.2 J.m$^{-2}$) [120] which are both higher than that of TiO$_2$ (1.78 J.m$^{-2}$) [121, 122]. This supports that Pt has a lower tendency to be encapsulated than Ru. The stronger tendency of Ru to oxidise compared to Pt also could play a role. Experimental evidence supporting this has been shown by Galhenage *et al.* [123], where bimetallic Pt-Ru clusters were grown on HOPG surfaces by vapor deposition. In this study, approximately even coverages of Pt and Ru were deposited, and after heating to 403 K the surface composition increased to 98–99% Pt, indicating that Pt has a greater surface affinity[123]. Another possible encapsulation mechanism may be that the Ru$_3$ clusters dissociate on the surface due to the interaction with the defect-rich titania, promoting encapsulation because less TiO$_2$ must be displaced by encapsulating a single atom, and single atoms are typically more mobile on oxide supports than small clusters [124]. This is possible given Ru-Ru has a bond dissociation energy of 193.0 ± 19.3 kJ/mol [125], which is lower than Pt-Pt, 307 ± 2 kJ/mol [126, 127].

*Table 4: Summary of findings from XPS, UPS and MIES.*

| Sample | XPS | UPS | MIES |
|---|---|---|---|
| Defect-TiO$_2$<br><br>heated to 723 K | Increasing Ti$^{3+}$ with Ar$^+$ sputtering | Modified O 2p states due to surface defects. Small amount of Ti$^{3+}$ features. | Ti defects, mainly Ti$^{3+}$. |
| SD-Ru3<br><br>heated to 723 K | BE shift of -1.9 eV | Ru3 clusters agglomerated and encapsulated. | No additional states at the surface. |
| CVD-Ru3<br><br>heated to 723 K | Ru BE shift of -0.6 eV, Ligands removed | Ru3 clusters agglomerated and encapsulated. Ru exists as metal. | No additional states at the surface. Ru exists as metal. |
| CS-Pt3<br><br>heated to 573 K | Pt BE shift of -0.5eV | Pt$_3$ features at 4 – 5 eV. Pt$_3$ forms discrete states at the surface. Pt$_3$ is not encapsulated. | Pt$_3$ features at 4 – 5 eV. Pt$_3$ forms discrete states at the surface. Pt$_3$ is not encapsulated. |

**Implication of the DOS for photocatalysis**

An efficient photocatalyst requires states below the Fermi level to be occupied by holes to drive the oxygen evolution reaction (OER) in an overall photocatalytic water splitting reaction thus at a binding energy above the valence band edge of the photocatalyst.[128] The presence of such states can be investigated with valence electron spectroscopy such as UPS and MIES. For the hydrogen evolution reaction (HER) we need unoccupied states below the conduction band edge.[128] The electron spectroscopy methods applied here are not suitable to investigate these states.

It was found here that $Pt_3$ clusters form discrete states at a binding energy above the VB edge of $TiO_2$ as can be seen in Figure 7. These states should be suitable for supporting the HER reaction. For using $Pt_3$ cluster as co-catalysts in a real photocatalyst, it would be required to explore a different pathway to deposit the $Pt_3$ clusters onto particulate photocatalyst. Cluster generated with a cluster source can in most cases deposit only line of sight thus it is difficult to use this deposition method for particulate based photocatalysts.

The Ru clusters deposited were found to form nanoparticles with metallic states as can be seen from the broad band starting just below Fermi level as can be seen in Figure 7. However, as Figure 7 also shows, the DOS formed by Ru covers a distinct region with the DOS almost vanishing around 4 eV. This range of DOS is separated from higher binding energy DOS of the Ru and thus the Ru nanoparticles formed potentially could be useful for creating OER co-catalysts. It has been demonstrated by Teramura, the deposition of $Ru_3(CO)_{12}$ onto a oxynitride including heating to temperatures similar to those applied in the present work resulting in the improvement of the photocatalyst formed by the oxynitride $(Ga_{1-x}Zn_x)-(N_{1-x}O_x)$.[129] The Ru nanoparticles formed could potentially also be beneficial for other oxide or oxynitride photocatalysts.

## Conclusions

In this study XPS, UPS, and MIES were performed on samples after heat treatment, and reference spectra for UPS and MIES were determined due to $Ru_3$ (SD and CVD depositions) and $Pt_3$ (CS depositions). RF-sputter deposited $TiO_2$ treated with heating and $Ar^+$ sputtering was used as the substrate. For $Ru_3$ clusters, UPS measured the DOS for clusters while the MIES spectrum had contributions from defected titania atop of the clusters, due to encapsulation by the substrate. Comparing SD-$Ru_3$ to CVD-$Ru_3$, it was found that the method of depositing $Ru_3(CO)_{12}$ onto the surface does not have a significant effect on the DOS of the $Ru_3$ clusters after ligand removal by heat treatment. For $Pt_3$ clusters the UPS and MIES spectra were very similar, which suggests $Pt_3$ clusters were present on the topmost surface layer after heating to 523 K, which is most likely a benefit for photocatalysis applications. The fact that $Ru_3$ becomes encapsulated but $Pt_3$ does not is potentially due to an energetic benefit for Ru encapsulation.

The outcome of the present work is that on $TiO_2$ Pt clusters can modify the outermost layer by adding discrete energy levels on the surface, whereas Ru clusters become encapsulated just below the surface and contribute a broad distribution of energy levels close to the Fermi level. Pt clusters would be suitable for adding co-catalyst sites in the outermost layer of $TiO_2$ while Ru clusters are suitable when encapsulated co-catalyst sites are required.

## Acknowledgment.

The work is supported by the US Army project FA5209-16-R-0017. Part of the work is supported by the Australian Solar Thermal Research Institute (ASTRI), a project supported by the Australian Government, through the Australian Renewable Energy Agency (ARENA). The work at the University of Utah was partly supported by the Air Force Office of Scientific Research, under AFOSR grant FA9550-19-1-0261. The authors acknowledge the facilities, and the scientific and technical assistance of Microscopy Australia (ROR: 042mm0k03) enabled by NCRIS and the government of South Australia at Flinders Microscopy and Microanalysis (ROR: 04z91ja70), Flinders University (ROR: 01kpzv902).

# References

1. D. Lee, R. L. Donkers, G. Wang, A. S. Harper and R. W. Murray, *J. Am. Chem. Soc.*, 2004, **126**, 6193-6199.
2. J. Jung, H. Kim and Y.-K. Han, *Journal of the American Chemical Society*, 2011, **133**, 6090-6095.
3. G. Ramakrishna, O. Varnavski, J. Kim, D. Lee and T. Goodson, *Journal of the American Chemical Society*, 2008, **130**, 5032-5033.
4. M. Valden, X. Lai and D. W. Goodman, *Science*, 1998, **281**, 1647-1650.
5. C. Xu, X. Lai, G. Zajac and D. Goodman, *Phys. Rev. B*, 1997, **56**, 13464.
6. C. P. Joshi, M. S. Bootharaju and O. M. Bakr, *J. Phys. Chem. Lett.*, 2015, **6**, 3023-3035.
7. H. Li, L. Li and Y. Li, *Nanotechnol. Rev.*, 2013, **2**, 515-528.
8. A. Sanchez, S. Abbet, U. Heiz, W. D. Schneider, H. Hakkinen, R. N. Barnett and U. Landman, *J. Phys. Chem. A*, 1999, **103**, 9573.
9. H. Zhang, S. Zuo, M. Qiu, S. Wang, Y. Zhang, J. Zhang and X. W. D. Lou, *Sci. Adv.*, 2020, **6**, eabb9823.
10. Y. Negishi, Y. Matsuura, R. Tomizawa, W. Kurashige, Y. Niihori, T. Takayama, A. Iwase and A. Kudo, *J. Phys. Chem. C*, 2015, **119**, 11224-11232.
11. W. Kurashige, R. Kumazawa, D. Ishii, R. Hayashi, Y. Niihori, S. Hossain, L. V. Nair, T. Takayama, A. Iwase, S. Yamazoe, T. Tsukuda, A. Kudo and Y. Negishi, *J. Phys. Chem. C*, 2018, **122**, 13669-13681.
12. M. Chen and D. Goodman, *Catal. Today*, 2006, **111**, 22-33.
13. F. S. Roberts, M. D. Kane, E. T. Baxter and S. L. Anderson, *Phys. Chem. Chem. Phys.*, 2014, **16**, 26443-26457.
14. U. Heiz and E. Bullock, *J. Mater. Chem.*, 2004, **14**, 564-577.
15. H.-J. Freund and G. Pacchioni, *Chem. Soc. Rev.*, 2008, **37**, 2224-2242.
16. G. Pacchioni, L. Giordano and M. Baistrocchi, *Phys. Rev. Lett.*, 2005, **94**, 226104.
17. D. Ricci, A. Bongiorno, G. Pacchioni and U. Landman, *Phys. Rev. Lett.*, 2006, **97**, 036106.
18. U. Diebold, *Surf. Sci. Rep.*, 2003, **48**, 53-229.
19. H. S. Al Qahtani, K. Kimoto, T. Bennett, J. F. Alvino, G. G. Andersson, G. F. Metha, V. B. Golovko, T. Sasaki and T. Nakayama, *J. Chem. Phys.*, 2016, **144**, 114703.
20. R. Bennett, C. Pang, N. Perkins, R. Smith, P. Morrall, R. Kvon and M. Bowker, *J. Phys. Chem. B*, 2002, **106**, 4688-4696.
21. Q. Fu, T. Wagner, S. Olliges and H.-D. Carstanjen, *J. Phys. Chem. B*, 2005, **109**, 944-951.
22. X. Zhao, J. Hrbek and J. A. Rodriguez, *Surf. Sci.*, 2005, **575**, 115-124.
23. D. Meier, G. Rizzi, G. Granozzi, X. Lai and D. Goodman, *Langmuir*, 2002, **18**, 698-705.
24. A. Berkó, I. Ulrych and K. Prince, *J. Phys. Chem. B*, 1998, **102**, 3379-3386.
25. F. X. Xiao, S. F. Hung, J. Miao, H. Y. Wang, H. Yang and B. Liu, *Small*, 2015, **11**, 554-567.
26. I. X. Green, W. Tang, M. Neurock and J. T. Yates, *Science*, 2011, **333**, 736-739.
27. A. C. Reber, S. N. Khanna, F. S. Roberts and S. L. Anderson, *J. Phys. Chem. C*, 2016, **120**, 2126-2138.
28. H. S. Al Qahtani, G. F. Metha, R. B. Walsh, V. B. Golovko, G. G. Andersson and T. Nakayama, *J. Phys. Chem. C*, 2017, **121**, 10781-10789.
29. F. S. Roberts, S. L. Anderson, A. C. Reber and S. N. Khanna, *J. Phys. Chem. C*, 2015, **119**, 6033-6046.
30. J.-Y. Ruzicka, F. Abu Bakar, C. Hoeck, R. Adnan, C. McNicoll, T. Kemmitt, B. C. Cowie, G. F. Metha, G. G. Andersson and V. B. Golovko, *J. Phys. Chem. C*, 2015, **119**, 24465-24474.
31. K. Nakata, S. Sugawara, W. Kurashige, Y. Negishi, M. Nagata, S. Uchida, C. Terashima, T. Kondo, M. Yuasa and A. Fujishima, *Int. J. Photoenergy*, 2013, **2013**, 456583.
32. K. Katsiev, G. Harrison, Y. Al-Salik, G. Thornton and H. Idriss, *ACS Catal.*, 2019, **9**, 8294-8305.
33. P. López-Caballero, A. W. Hauser and M. a. Pilar de Lara-Castells, *J. Phys. Chem. C*, 2019, **123**, 23064-23074.
34. Q. Hao, ZhiqiangWang, T. Wang, Z. Ren, C. Zhou and X. Yang, *ACS Catal.*, 2019, **9**, 286-294.
35. T. Kawai and T. Sakata, *J. Chem. Soc., Chem. Commun.*, 1980, 694-695.
36. Z. Shu, Y. Cai, J. Ji, C. Tang, S. Yu, W. Zou and L. Dong, *Catalysts*, 2020, **10**, 1047.

37. E. Kowalska, H. Remita, C. Colbeau-Justin, J. Hupka and J. Belloni, *J. Phys. Chem. C*, 2008, **112**, 1124-1131.
38. W. Ouyang, M. J. Munoz-Batista, A. Kubacka, R. Luque and M. Fernández-García, *Appl. Catal. B*, 2018, **238**, 434-443.
39. G. L. Chiarello, D. Ferri and E. Selli, *J. Catal.*, 2011, **280**, 168-177.
40. J. Xu, T. Liu, J. Li, B. Li, Y. Liu, B. Zhang, D. Xiong, I. Amorim, W. Li and L. Liu, *Energy Environ. Sci.*, 2018, **11**, 1819-1827.
41. T. Sreethawong and S. Yoshikawa, *Catal. Commun.*, 2005, **6**, 661-668.
42. X. Li, W. Bi, L. Zhang, S. Tao, W. Chu, Q. Zhang, Y. Luo, C. Wu and Y. Xie, *Adv. Mater.*, 2016, **28**, 2427-2431.
43. C. Moreno-Castilla, M. A. Salas-Peregrín and F. J. López-Garzón, *J. Mol. Catal. A: Chem.*, 1995, **95**, 223-233.
44. P. Panagiotopoulou, *Appl. Catal. A*, 2017, **542**, 63-70.
45. R. Mutschler, E. Moioli and A. Züttel, *J. Catal.*, 2019, **375**, 193-201.
46. G. D. Weatherbee and C. H. Bartholomew, *J. Catal.*, 1984, **87**, 352-362.
47. K. Asakura and Y. Iwasawa, *J. Chem. Soc., Faraday Trans.*, 1990, **86**, 2657-2662.
48. F. Solymosi, A. Erdöhelyi and M. Kocsis, *J. Chem. Soc., Faraday Trans. 1*, 1981, **77**, 1003-1012.
49. W. R. Hastings, C. J. Cameron, M. J. Thomas and M. C. Baird, *Inorganic Chemistry*, 1988, **27**, 3024-3028.
50. C. S. Kellner and A. T. Bell, *J. Catal.*, 1982, **75**, 251-261.
51. R. D. Gonzalez and H. Miura, *J. Catal.*, 1982, **77**, 338-347.
52. E. Kikuchi, H. Nomura, M. Matsumoto and Y. Morita, *Appl. Catal.*, 1983, **7**, 1-9.
53. G. A. Rizzi, A. Magrin and G. Granozzi, *Phys. Chem. Chem. Phys.*, 1999, **1**, 709-711.
54. T. Cai, Z. Song, Z. Chang, G. Liu, J. Rodriguez and J. Hrbek, *Surf. Sci.*, 2003, **538**, 76-88.
55. F. Yang, S. Kundu, A. B. Vidal, J. Graciani, P. J. Ramírez, S. D. Senanayake, D. Stacchiola, J. Evans, P. Liu and J. F. Sanz, *Angew. Chem. Int. Ed.*, 2011, **50**, 10198-10202.
56. L. Howard-Fabretto, T. J. Gorey, G. Li, S. Tesana, G. F. Metha, S. L. Anderson and G. G. Andersson, *Nanoscale Advances*, 2021.
57. L. Howard-Fabretto, T. J. Gorey, G. Li, D. J. Osborn, S. Tesana, G. F. Metha, S. L. Anderson and G. G. Andersson, *Physical Chemistry Chemical Physics*, 2024, **26**, 19117-19129.
58. K. Asakura, K.-K. Bando and Y. Iwasawa, *J. Chem. Soc., Faraday Trans.*, 1990, **86**, 2645-2655.
59. T. Choudhary, C. Sivadinarayana, C. C. Chusuei, A. Datye, J. Fackler Jr and D. Goodman, *J. Catal.*, 2002, **207**, 247-255.
60. H. S. Al Qahtani, R. Higuchi, T. Sasaki, J. F. Alvino, G. F. Metha, V. B. Golovko, R. Adnan, G. G. Andersson and T. Nakayama, *RSC Adv.*, 2016, **6**, 110765-110774.
61. G. G. Andersson, V. B. Golovko, J. F. Alvino, T. Bennett, O. Wrede, S. M. Mejia, H. S. Al Qahtani, R. Adnan, N. Gunby and D. P. Anderson, *J. Chem. Phys.*, 2014, **141**, 014702.
62. H. Morgner, *Adv. At. Mol. Opt. Phys.*, 2000, **42**, 387-488.
63. M. Seah and W. Dench, *Surf. Interface Anal.*, 1979, **1**, 2-11.
64. W. Eberhardt, P. Fayet, D. Cox, Z. Fu, A. Kaldor, R. Sherwood and D. Sondericker, *Phys. Rev. Lett.*, 1990, **64**, 780.
65. J. Daughtry, A. Alotabi, L. Howard-Fabretto and G. G. Andersson, *Nanoscale Advances*, 2020, **3**, 1077-1086.
66. T. J. Gorey, B. Zandkarimi, G. Li, E. T. Baxter, A. N. Alexandrova and S. L. Anderson, *J. Phys. Chem. C*, 2019, **123**, 16194-16209.
67. G. Li, B. Zandkarimi, A. C. Cass, T. J. Gorey, B. J. Allen, A. N. Alexandrova and S. L. Anderson, *J. Chem. Phys.*, 2020, **152**, 024702.
68. T. J. Gorey, Y. Dai, S. L. Anderson, S. Lee, S. Lee, S. Seifert and R. E. Winans, *Surf. Sci.*, 2020, **691**, 121485.
69. E. Spiller, *Soft X-ray optics*, SPIE Optical Engineering Press Bellingham, WA, 1994.
70. Y. Harada, S. Masuda and H. Ozaki, *Chem. Rev.*, 1997, **97**, 1897-1952.
71. G. Krishnan, N. Eom, R. M. Kirk, V. B. Golovko, G. F. Metha and G. G. Andersson, *J. Phys. Chem. C*, 2019, **123**, 6642−6649.
72. G. Krishnan, H. S. Al Qahtani, J. Li, Y. Yin, N. Eom, V. B. Golovko, G. F. Metha and G. G. Andersson, *J. Phys. Chem. C*, 2017, **121**, 28007-28016.

73. R. Schlaf, P. Schroeder, M. Nelson, B. Parkinson, C. Merritt, L. Crisafulli, H. Murata and Z. Kafafi, *Surf. Sci.*, 2000, **450**, 142-152.
74. R. A. Bennett, J. Mulley, M. Newton and M. Surman, *J. Chem. Phys.*, 2007, **127**, 084707.
75. L. Zhang, R. Persaud and T. E. Madey, *Phys. Rev. B*, 1997, **56**, 10549.
76. T. Bennett, R. H. Adnan, J. F. Alvino, V. Golovko, G. G. Andersson and G. F. Metha, *Inorganic Chemistry*, 2014, **53**, 4340-4349.
77. H. Morgner, in *Advances In Atomic, Molecular, and Optical Physics*, eds. B. Benjamin and W. Herbert, Academic Press, 2000, vol. Volume 42, pp. 387-488.
78. C. C. Chusuei, X. Lai, K. Luo and D. Goodman, *Topics in Catalysis*, 2000, **14**, 71-83.
79. W. E. Kaden, T. Wu, W. A. Kunkel and S. L. Anderson, *Science*, 2009, **326**, 826-829.
80. N. Isomura, X. Wu, H. Hirata and Y. Watanabe, *J. Vac. Sci. Technol. A*, 2010, **28**, 1141-1144.
81. Y. Dai, T. J. Gorey, S. L. Anderson, S. Lee, S. Lee, S. Seifert and R. E. Winans, *J. Phys. Chem. C*, 2017, **121**, 361-374.
82. W. E. Kaden, C. Büchner, L. Lichtenstein, S. Stuckenholz, F. Ringleb, M. Heyde, M. Sterrer, H.-J. Freund, L. Giordano and G. Pacchioni, *Phys. Rev. B*, 2014, **89**, 115436.
83. M. D. Kane, F. S. Roberts and S. L. Anderson, *Int. J. Mass Spectrom. Ion Processes*, 2014, **370**, 1-15.
84. C. N. Wagner, AV; Kraut-Vass, A; Allison, JW; Powell, CJ; Rumble, JR, Jr., *Journal*, 2002.
85. G. B. Fisher, *Surf. Sci.*, 1979, **87**, 215-227.
86. J. Kim, J. Chung and S.-J. Oh, *Phys. Rev. B*, 2005, **71**, 121406.
87. J. Okamoto, S.-I. Fujimori, T. Okane, A. Fujimori, M. Abbate, S. Yoshii and M. Sato, *Phys. Rev. B*, 2006, **73**, 035127.
88. K. Schierbaum, S. Fischer, M. Torquemada, J. De Segovia, E. Roman and J. Martin-Gago, *Surf. Sci.*, 1996, **345**, 261-273.
89. J. Crowell, E. Garfunkel and G. Somorjai, *Surf. Sci.*, 1982, **121**, 303-320.
90. E. I. Altman and R. J. Gorte, *Surface Science*, 1989, **216**, 386-394.
91. S. Krischok, J. Schaefer, O. Höfft and V. Kempter, *Surf. Interface Anal.*, 2005, **37**, 83-89.
92. S. Krischok, J. Günster, D. Goodman, O. Höfft and V. Kempter, *Surf. Interface Anal.*, 2005, **37**, 77-82.
93. S. Pan, X. Liu, M. Guo, S. fung Yu, H. Huang, H. Fan and G. Li, *J. Mater. Chem. A*, 2015, **3**, 11437-11443.
94. S. Labich, E. Taglauer and H. Knözinger, *Topics in Catalysis*, 2000, **14**, 153-161.
95. F. Pesty, H.-P. Steinrück and T. E. Madey, *Surf. Sci.*, 1995, **339**, 83-95.
96. Y. Gao, Y. Liang and S. Chambers, *Surf. Sci.*, 1996, **365**, 638-648.
97. H. R. Sadeghi and V. E. Henrich, *J. Catal.*, 1988, **109**, 1-11.
98. H. R. Sadeghi and V. E. Henrich, *Applications of Surface Science*, 1984, **19**, 330-340.
99. O. Dulub, W. Hebenstreit and U. Diebold, *Phys. Rev. Lett.*, 2000, **84**, 3646.
100. T. Komaya, A. T. Bell, Z. Wengsieh, R. Gronsky, F. Engelke, T. S. King and M. Pruski, *J. Catal.*, 1994, **149**, 142-148.
101. S. Bernal, F. Botana, J. Calvino, C. López, J. Pérez-Omil and J. Rodríguez-Izquierdo, *J. Chem. Soc., Faraday Trans.*, 1996, **92**, 2799-2809.
102. A. Wang, J. Li and T. Zhang, *Nature Reviews Chemistry*, 2018, **2**, 65-81.
103. O. Dulub, W. Hebenstreit and U. Diebold, *Physical Review Letters*, 2000, **84**, 3646-3649.
104. C. Mead and W. Spitzer, *Phys. Rev. Lett.*, 1963, **10**, 471.
105. G. Wertheim and S. DiCenzo, *Phys. Rev. B*, 1988, **37**, 844.
106. S. DiCenzo and G. Wertheim, *Solid State Phys.*, 1985, **11**, 203.
107. D. J. Alberas, J. Kiss, Z.-M. Liu and J. M. White, *Surf. Sci.*, 1992, **278**, 51-61.
108. C. Vayenas, S. Brosda and C. Pliangos, *J. Catal.*, 2003, **216**, 487-504.
109. T. Ioannides and X. E. Verykios, *J. Catal.*, 1996, **161**, 560-569.
110. S. Tauster, S. Fung and R. L. Garten, *J. Am. Chem. Soc.*, 1978, **100**, 170-175.
111. S. Tauster, S. Fung, R. Baker and J. Horsley, *Science*, 1981, **211**, 1121-1125.
112. G. L. Haller and D. E. Resasco, in *Advances in catalysis*, Elsevier, 1989, vol. 36, pp. 173-235.
113. R. Bennett, P. Stone and M. Bowker, *Catal. Lett.*, 1999, **59**, 99-105.
114. D. Mullins and K. Zhang, *Surf. Sci.*, 2002, **513**, 163-173.
115. H.-V. Roy, P. Fayet, F. Patthey, W.-D. Schneider, B. Delley and C. Massobrio, *Phys. Rev. B*, 1994, **49**, 5611.

116. V. Çelik, H. Ünal, E. Mete and Ş. Ellialtıoğlu, *Phys. Rev. B*, 2010, **82**, 205113.
117. Z. Wu, Y. Li and W. Huang, *J. Phys. Chem. Lett.*, 2020.
118. M. Macino, A. J. Barnes, S. M. Althahban, R. Qu, E. K. Gibson, D. J. Morgan, S. J. Freakley, N. Dimitratos, C. J. Kiely and X. Gao, *Nat. Catal.*, 2019, **2**, 873-881.
119. Q. Fu and T. Wagner, *J. Phys. Chem. B*, 2005, **109**, 11697-11705.
120. W. Tyson and W. Miller, *Surface Science*, 1977, **62**, 267-276.
121. M. Aizawa, S. Lee and S. L. Anderson, *Surf. Sci.*, 2003, **542**, 253-275.
122. A. Howard, C. Mitchell, D. Morris, R. Egdell and S. Parker, *Surf. Sci.*, 2000, **448**, 131-141.
123. R. P. Galhenage, K. Xie, W. Diao, J. M. M. Tengco, G. S. Seuser, J. R. Monnier and D. A. Chen, *Phys. Chem. Chem. Phys.*, 2015, **17**, 28354-28363.
124. C. T. Campbell, *Acc. Chem. Res.*, 2013, **46**, 1712-1719.
125. Y.-R. Luo, *Comprehensive handbook of chemical bond energies*, CRC press, Boca Raton, FL, 2007.
126. D. R. Lide, *CRC handbook of chemistry and physics: a ready-reference book of chemical and physical data*, CRC press, 1995.
127. S. Taylor, G. W. Lemire, Y. M. Hamrick, Z. Fu and M. D. Morse, *J. Chem. Phys.*, 1988, **89**, 5517-5523.
128. Q. Wang and K. Domen, *Chem. Rev.*, 2020, **120**, 919-985.
129. K. Teramura, K. Maeda, T. Saito, T. Takata, N. Saito, Y. Inoue and K. Domen, *J. Phys. Chem. B*, 2005, **109**, 21915-21921.

# Supplementary Material: Density of States of Ru$_3$ and Pt$_3$ clusters supported on Sputter-Deposited TiO$_2$

Liam Howard-Fabretto[1,2], Timothy J. Gorey[3], Guangjing Li[3], Siriluck Tesana[4], Gregory F. Metha[5], Scott L. Anderson[3], and Gunther G. Andersson[1,2]*

1 Flinders Institute for Nanoscale Science and Technology, Flinders University, Adelaide, South Australia 5042, Australia

2 Flinders Microscopy and Microanalysis, College of Science and Engineering, Flinders University, Adelaide, South Australia 5042, Australia

3 Chemistry Department, University of Utah, 315 S. 1400 E., Salt Lake City, UT 84112, United States

4 The MacDiarmid Institute for Advanced Materials and Nanotechnology, School of Physical and Chemical Sciences, University of Canterbury, Christchurch 8141, New Zealand

5 Department of Chemistry, University of Adelaide, Adelaide, South Australia 5005, Australia

*Corresponding author: Gunther G. Andersson

Email: gunther.andersson@flinders.edu.au.

Address: Physical Sciences Building (2111) GPO Box 2100, Adelaide 5001, South Australia

# Background

## MIES Background

In a UPS measurement, incident ultraviolet (UV) light causes outer valence electrons to be ejected from the sample as photoelectrons due to the photoelectric effect. UPS can be used to measure the electronic properties of a surface such as the occupied density of states (DOS) [1], featuring information depth of ~2-3 nm [2]. Metastable impact electron spectroscopy (MIES) can also be used to measure surface electronic properties such as DOS [3] in a manner similar to UPS. However, rather than firing photons at the surface, metastable helium atoms (He*) with an excitation of 19.8 eV [3] are used which bombard the surface, resulting in a He* de-excitation or neutralisation process and the emission and detection of an electron.

The major advantage of MIES is that it is purely surface sensitive because He* are not able to exceed the Van der Waals interactions at the surface [3, 4], unlike UPS which has an information depth of 2-3 nm [2]. When used together these techniques are complimentary; as an example, a previous study by Chambers *et al.* [5] used UPS and MIES to analyse the valence structure of double-walled carbon nanotubes, and it was shown that UPS measures the DOS across the whole nanotube while MIES only measured the DOS of the outer nanotube layer. However, previous studies using MIES on supported metal clusters are limited. Several MIES studies by Andersson *et al.* [6-8] have been performed on small, $TiO_2$-supported Au clusters, where the surface sensitivity of MIES was utilised for measuring clusters on the topmost layer.

The mechanism by which an electron is emitted from the surface in MIES is not unique. Electron emissions can occur by two known de-excitation pathways; these are 1) resonance ionisation (RI) followed by Auger neutralisation (AN), and 2) Auger de-excitation (AD) [4, 9]. This has been previously described in detail by Morgner [4]. Any MIES measurement may be some combination of AD and RI+AN to varying degrees depending upon the composition and temperature of the surface [4]. RI is hindered by surfaces with an absence of unoccupied states which the excited He* electron can tunnel to [3]. Thus, typically RI+AN is supressed for non-metallic samples, while it dominates for high work function metals. Conversely, the AD process dominates for semiconducting, organic, and insulating surfaces, as well as surfaces with very low work functions [4, 10]. It is possible that a surface could have multiple unoccupied surface states, which leads to features being broadened in the RI+AN spectra [3, 4]. The resulting spectra are directly comparable between UPS and the AD components of MIES, where both show the same features but the relative sizes may be different [4].

# Experimental

## Cluster Source Depositions of Pt

A 99.9% pure Pt target was vaporized by pulsed laser vaporization. Prior to CS depositions, substrates were liquid $N_2$ cooled to 180 K, followed by a quick flash to 700 K to remove any adventitious hydrocarbons. As the samples cooled after flash heating, deposition was started at 300 K and continued as the samples to 180 K. A RF quadrupole ion guide and bending unit were set to collect positively charged clusters. The beam passed through several differential pumping stages, before reaching a quadrupole mass filter which selected for a specific mass/charge ratio. Because multiply charged clusters have negligible intensity in the source distribution,[11, 12] this process selects a particular cluster size selection. Following size-selection, the clusters where guided into the UHV system, and passed through a 2 mm diameter aperture positioned 0.5 mm from the substrate surface, which produces cluster spots of ~2 mm in diameter. A retarding potential was used to achieve a deposition energy of ~1 eV/atom to prevent cluster damage or embedding [13]. Because the electron spectrometer analysis area was substantially larger than 2 mm, each CS-$Pt_3$ sample had 7 individual $Pt_3$ cluster spots deposited such that the spots did not overlap. Two CS-$Pt_3$/$TiO_2$ samples were prepared which had coverages of nominally 3 x $10^{13}$ atoms/$cm^2$ and 4.5 x $10^{13}$ atoms/$cm^2$ based on the measured neutralisation current during depositions. These values only include the cluster-deposited areas, not the blank substrate area in between. The ratio of cluster-covered area in the electron spectroscopy analysis area was ~0.22. A higher number of repeat samples for the CS-$Pt_3$ series would have been ideal for consistency with the other series, but in this case the number of samples was found to be sufficient for determining the UPS and MIES reference spectra due to the $Pt_3$ clusters.

**XPS**

**Peak Fitting**

CasaXPS was used to fit the peaks in the XPS spectra. Shirley backgrounds were subtracted from each spectrum before peak fitting. Individual reference peaks were typically fitted to the measured spectra using a convolution of Gaussian and Lorentzian line shapes. Peaks were fitted based on area, full width half maximum (FWHM) and peak location. The uncertainty in measured BEs is ± 0.2 eV, however in this study the presented BE's are averages from multiple samples and it was determined the weighted average standard deviation (taken to be the uncertainty) for cluster peak locations (for Ru 3d and Pt 4f) was ± 0.1 eV.

Every XPS spectrum featured C 1s peaks to some extent. For every sample peaks assigned to C-C or C-H were present at 285.0 eV, and C=O or C-O-C were present at 287.0 eV. A third carbon peak at 289.4 eV, assigned to O=C-O, was sometimes present but was typically removed by heat treatment and sputtering. These results are comparable to previously reported assignments for carbon contamination on $SiO_2$ substrates [14].

In the Ru 3d region, the adventitious C 1s peak overlaps with the Ru 3d doublet for clusters. To help with fitting the Ru 3d peak, a metallic Ru reference metal was analysed with XPS in our previous publication, and used as a fitting model for cluster Ru 3d spectra [15]. When comparing the $3d_{5/2}$ and $3d_{3/2}$ peaks, the peak separation was 4.17 eV, peak area ratio was 3:2, and FWHM ratio was 1:1.15.

Ru typically features asymmetrical line shapes for the 3d core electrons, and studies have been performed by Morgan [16] investigating the best way to fit this. The extent of asymmetry is dependent on the chemical nature of Ru as well as the resolution of the XPS instrument [16-18]. The line shapes published by Morgan were used as a starting point and were altered to best fit the line shape seen for the Ru clusters; LF(0.75,1.25,500,250) for Ru $3d_{5/2}$ and LF(0.8,1,500,250) for Ru $3d_{3/2}$. LF indicates Lorentzian asymmetric line shapes with tail damping.

For $Pt_3$ clusters the Pt 4f peaks were symmetrical. The Pt 4f region was fitted with 3 peaks, related to the Pt 4f doublet as well as an energy loss (E loss) peak in the same region which was also present for blank $TiO_2$. The fitting results for the blank $TiO_2$ were used as a template for the peak location and full width half maximum (FWHM) of the E loss peak when fitting results for samples with CS-$Pt_3$ present.

**Calculation of Surface Coverage**

Atomic concentrations as a percentage (At%) were determined using XPS. All the peaks associated with elements present on the surface were fitted and integrated to determine their peak area, and the areas were calibrated by dividing by XPS sensitivity factors found in the Handbook of X-ray Photoelectron Spectroscopy [19]. These were C 1s = 0.296, O 1s = 0.711, Ti 2p = 2.001, Ru 3d = 4.273, and Pt 4f = 5.575. The atomic ratios were determined by dividing the calibrated peak area for each element by the total calibrated peak area for all elements, and these were multiplied by 100% to determine the At%. The ratio of $Ti^{3+}$ and $Ti^{2+}$ peaks to the total summed Ti peak signals ($Ti^{Defect}/Ti^{Total}$, or Ti defect ratio) was calculated for each Defect-$TiO_2$ sample, and is used as a measurement for the level of defects in the titania.

The surface coverage of clusters was calculated for each sample and is presented in units of % ML, which is the percentage of the surface covered relative to a monolayer (ML). This calculation was done using a procedure similar to Eschen *et al.* [20], except for the difference that Eschen *et al.* used multiple XPS detection angles. This process solved for the surface coverage required to achieve the measured XPS At% for each cluster metal (Ru or Pt). The clusters were assumed to be present in only a single ML on the surface with negligible stacking of atoms and no mixing of cluster and substrate layers. The bulk interatomic distance for each metal was used to estimate the layer thickness for deposited Ru and Pt clusters, where Ru-Ru is 0.265 nm and Pt-Pt is 0.278 nm [21]. The contribution of individual atoms to the XPS spectra is reduced as the depth of the atom into the surface increases, which was also factored into the calculation by using the IMFP of electrons in $TiO_2$, 1.8 nm [22]. For $Pt_3$ measurements a scaling factor based on the ratio of the analysed surface covered by clusters was used in calculations.

**Uncertainties**

The relative uncertainty in At% varied depending on the size of the XPS peaks, and whether they overlapped with other peaks. Due to this, the uncertainty in At% for each peak of interest was estimated by determining the largest range of peak areas which resulted in what was considered a reasonable peak fitting. The relative uncertainty for At% of deposited cluster peaks was ~4% for both Ru 3d and Pt 4f. For elements in the substrate, the relative uncertainty is ~1% for $Ti^{4+}$ 2p and O 1s in $TiO_2$. The defects in $TiO_2$ spectra are present as $Ti^{3+}$ and $Ti^{2+}$ 2p peaks (these, when summed, are referred to as $Ti^{defects}$), and each has a relative uncertainty of ~15%.

A range of factors contribute to the uncertainty in surface coverage. These include errors in the calculated At% for the clusters, differences between atomic sensitivity factors in our detector setup and in the XPS handbook [19], and any inaccuracy in the IMFP of electrons in the substrate. Based mainly on the uncertainty of the IMFP, the absolute error in surface coverages was assumed to be ~100%. However, the relative error comparing between samples is based only on the cluster At% uncertainty and is ~4%. While the ~100% error can be considered high, the surface coverage estimation was intended only to give the scale of the surface coverage of clusters used in the experiments.

**UPS and MIES**

**Determining Reference Spectra**

To determine the reference spectrum due to deposited metal clusters, a blank $TiO_2$ spectrum was subtracted from each spectrum of cluster-loaded samples which were multiplied by linear scaling factors. The scaling factors were determined by minimising the differences between the resultant reference spectra for all samples in a series using the Excel Solver function. The resultant reference spectra from each sample were then averaged to determine an average reference spectrum for the sample series. In this process an assumption is made that the measured spectra are a linear combination of the substrate spectrum, and a spectrum related to the clusters. This assumes that the surface coverage does not affect the DOS of the clusters, which is reasonable considering the cluster-loaded surfaces were <20% ML, and similar assumptions for MIES spectra have been found to hold true for several systems [4]. The reference spectra were in some cases multiplied by a scaling factor for the figures for ease of comparison between spectra. For each sample series, weighting factors were determined for the contributions of the substrate and clusters to each measured spectrum, which provide evidence that the determined reference spectra are related to the cluster depositions. This is discussed further in the Supplementary Material.

**Weighting Factors**

Weighting factors for the substrate and cluster reference spectra from UPS and MIES were determined by linearly combining the substrate spectrum and the average reference spectrum. Weighting factors were determined using the Excel Solver function to minimise the sum of differences between the measured spectrum and the linear combination of reference spectra. For CS-$Pt_3$ the ratio of the substrate which was covered in clusters was factored into the reported weighting factors. The weighting factors were then normalised so that their sum was unity. The uncertainty in weighting factors was estimated to be ± 0.05.

# Results

## XPS

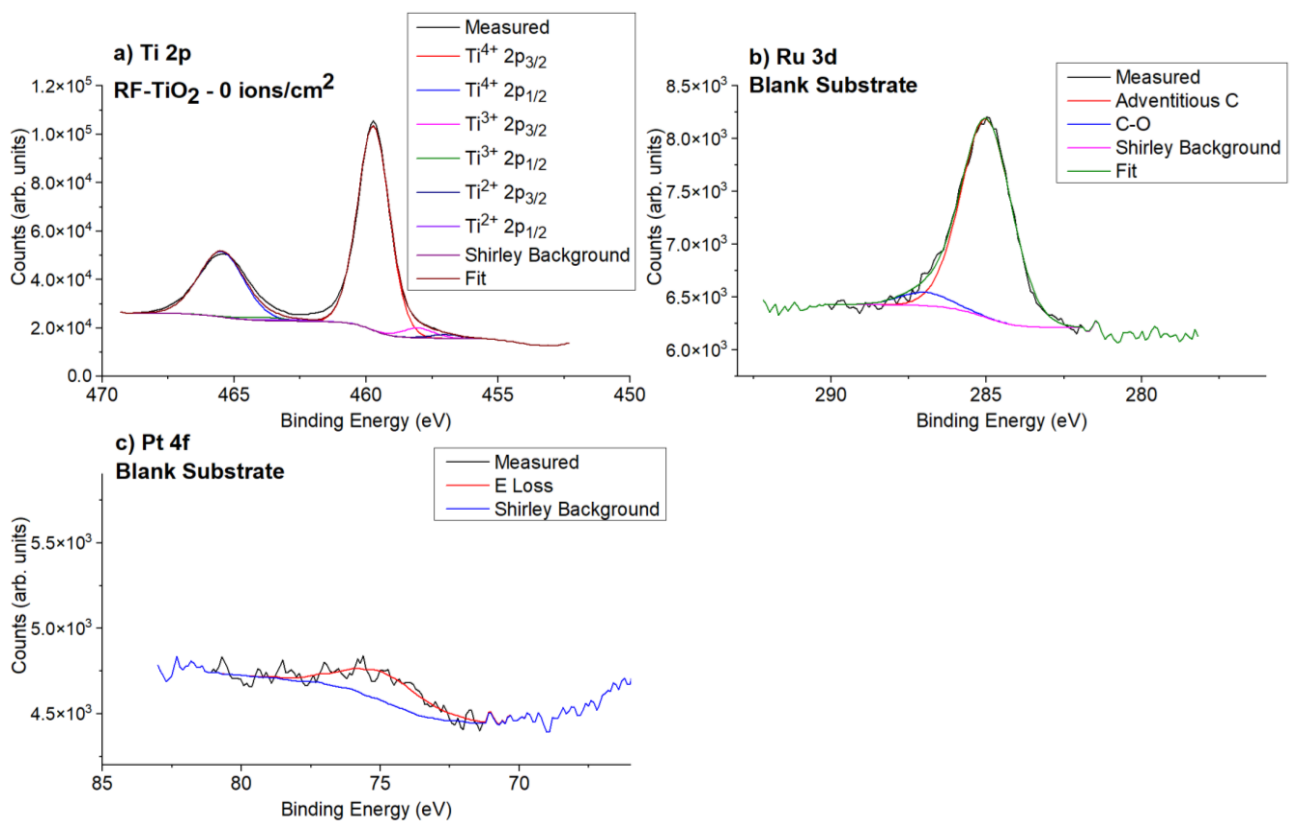

*Figure S1: XPS peak fitting examples for blank samples with no clusters, which were used to assist with the fitting of the cluster-loaded samples. a) Ti 2p peak fitting for TiO$_2$, after heating to 723 K but prior to treatment with Ar$^+$ sputtering. b) Ru 3d/C 1s peak fitting for heated and sputtered TiO$_2$ prior to Ru cluster deposition. c) Pt 4f region fitting for heated and sputtered TiO$_2$ with no deposited Pt clusters.*

Table S1: XPS results for each UPS/MIES sample after heating to 723 K. The At% was calculated for species of interest and is shown for $Ti^{4+}$, $(Ti^{3+} + Ti^{2+})$, and Ru or Pt. See the Methodology of the Supplementary Material for information on the relative uncertainties for At%.

| Sample Series | Sample | $Ti^{4+}$ At% | $(Ti^{3+} + Ti^{2+})$ At% | Ru or Pt At% |
|---|---|---|---|---|
| **SD-Ru$_3$** | SD-Blank | 24.5 | 1.6 | 0 |
|  | SD-Ru$_3$-1 | 23.8 | 1.5 | 0.04 |
|  | SD-Ru$_3$-2 | 24.6 | 1.7 | 0.1 |
|  | SD-Ru$_3$-3 | 23.4 | 0.8 | 0.5 |
|  | SD-Ru$_3$-4 | 24.3 | 1.7 | 0.8 |
|  | SD-Ru$_3$-5 | 22.9 | 1.8 | 1.2 |
|  | SD-Ru$_3$-6 | 22.1 | 1.7 | 1.3 |
| **CVD-Ru$_3$** | CVD-Blank | 24.8 | 1.7 | 0 |
|  | CVD-Ru$_3$-1 | 23.9 | 2.0 | 0.2 |
|  | CVD-Ru$_3$-2 | 24.3 | 1.9 | 0.2 |
|  | CVD-Ru$_3$-3 | 23.0 | 2.2 | 0.4 |
|  | CVD-Ru$_3$-4 | 22.5 | 2.9 | 0.5 |
|  | CVD-Ru$_3$-5 | 23.7 | 2.2 | 0.6 |
| **CS-Pt$_3$** | CS-Blank | 23.2 | 1.9 | 0 |
|  | CS-Pt-1 | 22.5 | 1.8 | 0.07 |
|  | CS-Pt-2 | 22.4 | 2.2 | 0.09 |

**UPS/MIES**

**Sample Series Weighting Factors**

The weighting factors for the contributions of the blank substrate and the cluster reference spectra (or Ti defects) are shown in Figure S2. If the weighting factor for a spectrum related to clusters increases as the cluster surface coverage increases, this provides evidence that the determined reference spectrum is indeed related to the presence of the clusters.

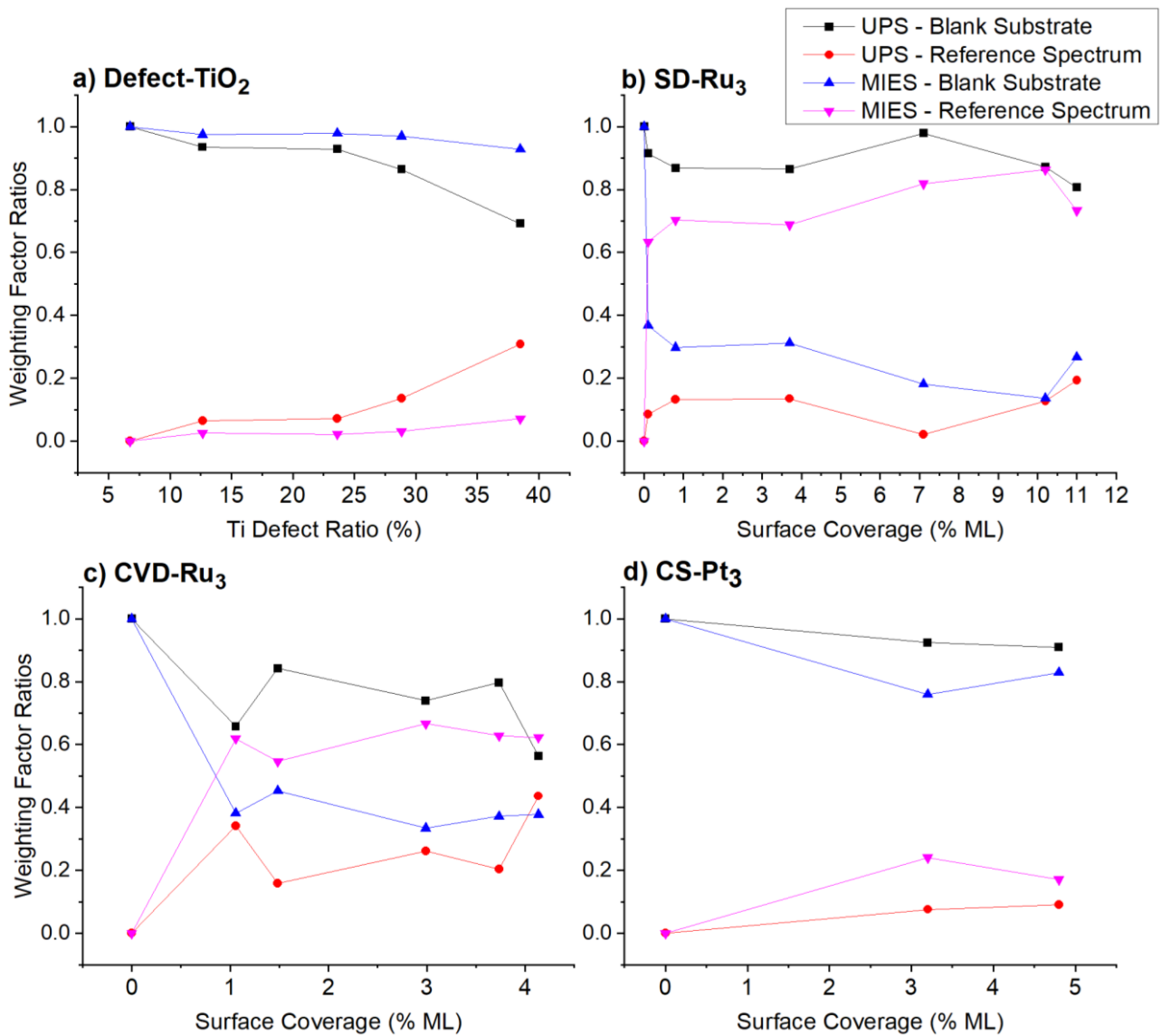

*Figure S2: UPS/MIES weighting factors for the blank substrate spectrum and the calculated reference spectrum for each sample series. a) Defect-TiO$_2$. b) SD-Ru$_3$. c) CVD-Ru$_3$. d) CS-Pt$_3$. Weighting factors are plotted as a function of the Ti defect ratio (a) or cluster surface coverage (b-d) and have an uncertainty of ± 0.05.*

The weighting factors for the Defect-TiO$_2$ reference spectra in both UPS and MIES increase approximately linearly with Ti$^{Defect}$/Ti$^{Total}$ (Figure S2a). This provides evidence that the determined UPS/MIES reference spectra for Defect-TiO$_2$ are related to the Ar$^+$ sputter treatment, and thus most likely represent the UPS/MIES signal for surface defects.

The SD-Ru$_3$ UPS and MIES weighting factors increase approximately with Ru surface coverage (Figure S2b), providing evidence that the SD-Ru$_3$ reference spectra are related to the cluster depositions. The weighting factor for UPS increases with Ru surface coverage as expected, except for the 7.1% ML sample which was lower than the trend and appears to be an outlier. The MIES weighting factor for SD-Ru$_3$ increased from 0 to 0.63 for the lowest surface concentration compared to the blank, and then increases further with surface concentration. The large initial increase is most likely due to the high surface sensitivity of MIES, where the surface layer is being changed more dramatically than for UPS due to the presence of the clusters.

The weighting factor for the UPS CVD-$Ru_3$ reference spectrum increases approximately linearly with Ru surface coverage (Figure S2c). The 1.1% ML measurement sample (1 minute CVD deposition) appears to be an outlier with a higher than expected weighting factor. The MIES weighting factor ratios for the CVD-$Ru_3$ reference spectrum increased from 0 to 0.62 for the lowest surface concentration, and then increases slightly further with surface coverage at higher concentrations, similar to what was seen for the MIES of SD-$Ru_3$. The increase in weighting factors is sufficient evidence that the determined UPS and MIES spectra for CVD-$Ru_3$ are related to the CVD depositions.

The CS-$Pt_3$ UPS reference spectrum weighting factor increases approximately linearly with Pt surface coverage, which provides evidence the CS-$Pt_3$ UPS reference spectrum is related to the deposition of $Pt_3$ clusters. The MIES weighting factor ratio for CS-$Pt_3$ increases from the blank (0% loading) to the 3.2% ML sample, but then decreases slightly for the 4.8% ML sample. However, this is deemed acceptable within the ± 0.05 uncertainty for the weighting factor ratios, and it is most likely the CS-$Pt_3$ MIES reference spectrum is also related to the deposition of $Pt_3$ clusters.

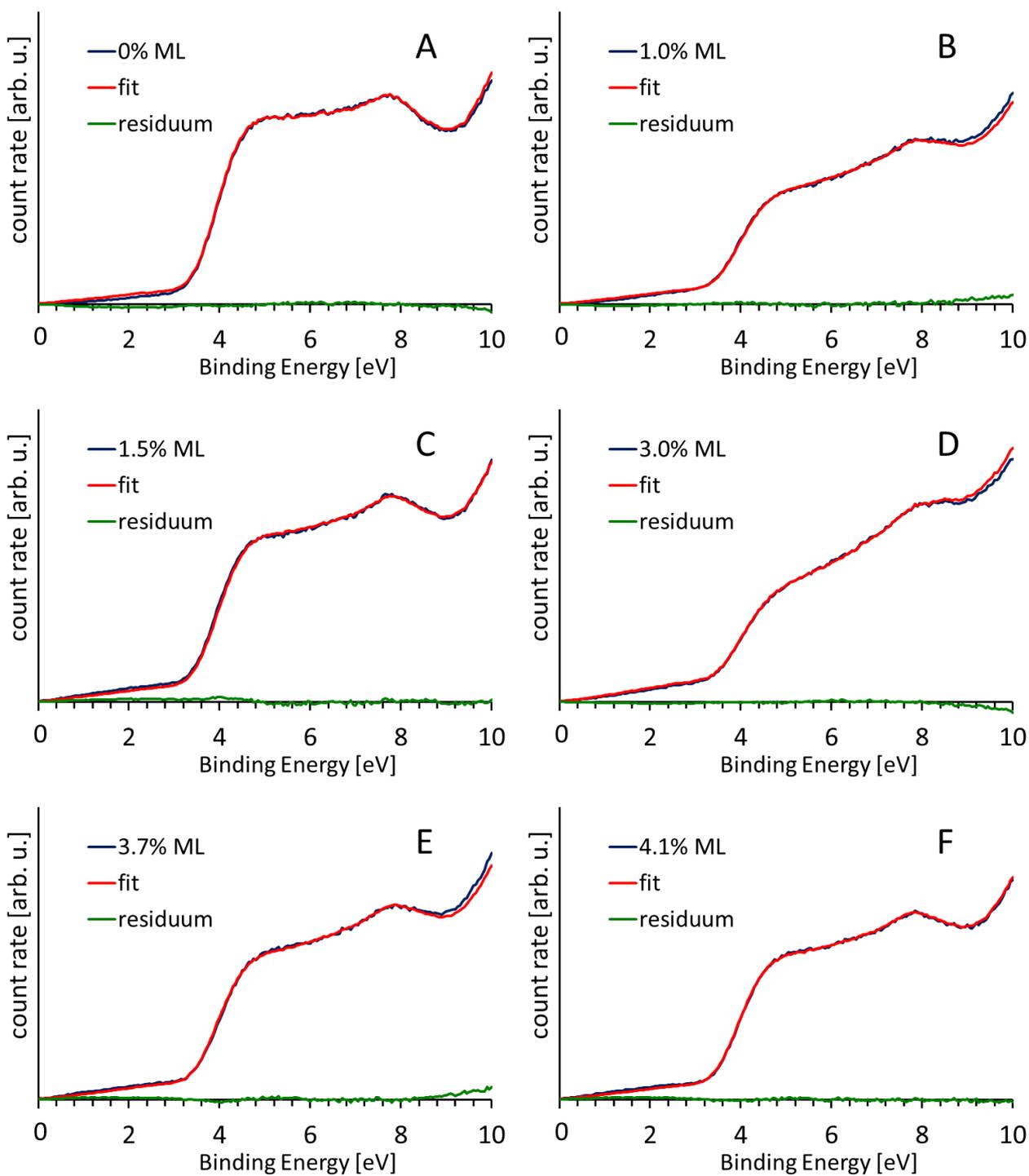

Figure S3: fits of the UPS results for the CVD-Ru3 sample series after heating for the loadings investigated (A – F).

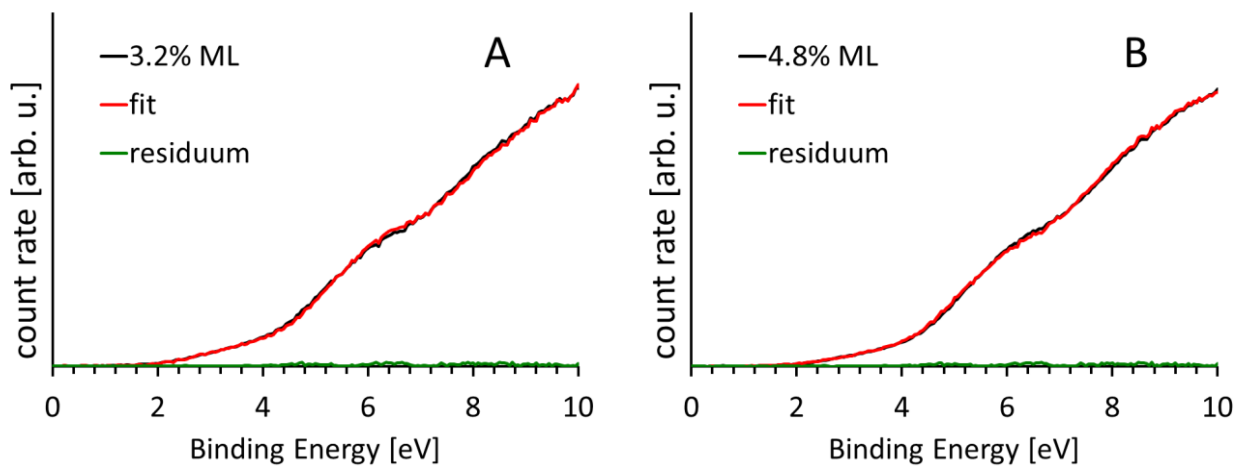

*Figure S4: fits of the MIES results for the CS-Pt3 sample series after heating for the loadings investigated (A – B).*

**Metallic Reference Sample**

A UPS/MIES measurement was performed on a 99.9% pure sample of bulk, metallic Ru, and is shown in Figure S5. The sample was treated by heating to 1073 K for 10 minutes and sputtering with 3 keV $Ar^+$ ions for 1 hour to remove the surface Ru oxide layer and any hydrocarbon contamination, which was confirmed by XPS.

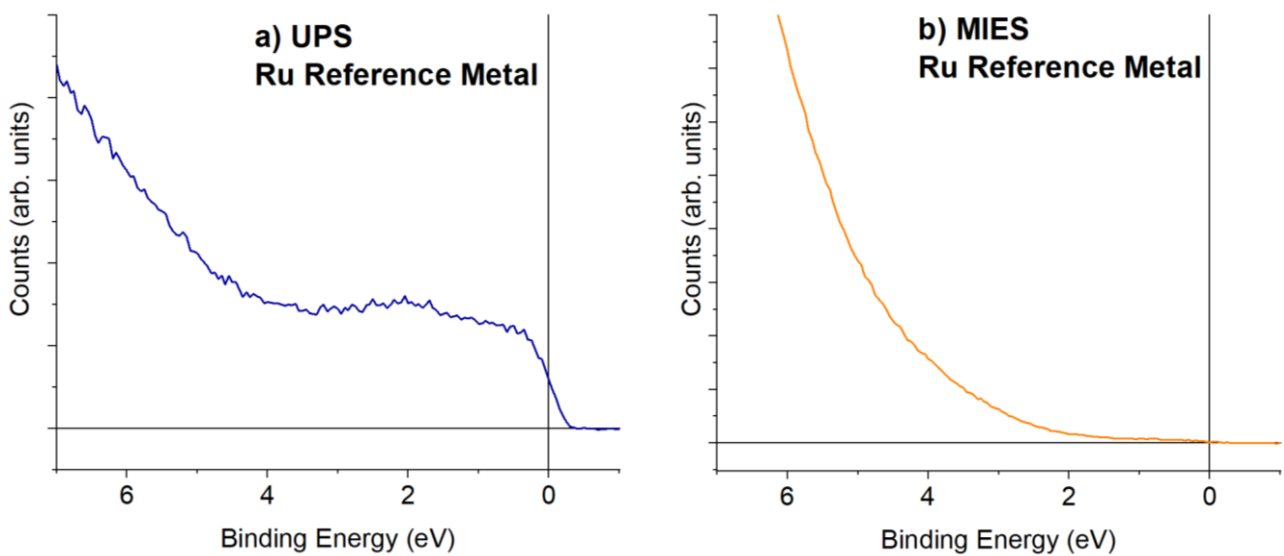

*Figure S5: Valence electron spectra for bulk, metallic Ru. a) UPS results, b) MIES results.*

The UPS spectrum for metallic Ru (Figure S5a) shows an onset at the Fermi level, which is expected for a bulk metallic sample [23-27], and has features present at 0.3 eV and 2.0 eV. The MIES spectrum (Figure S5b) has no distinct features, and features a broad, increasing background. This is because the MIES spectrum was the result of Auger neutralisation (AN) and resonance ionisation (RI), which is typical for metallic samples and results in a spectrum which is broadened compared to the true valence DOS [4]. Note that the BE axis of Figure S5b does not strictly show the true BE with regards to the DOS due to this broadening, but the BE axis was still used for consistency.

A bulk, high purity metallic Pt reference material was not available for analysis with UPS and MIES in this study. However, a reference UPS measurement of the Pt(111) surface is available in a study by Crowell *et al*. [28], where the onset of the spectrum was at the fermi level as per the Ru UPS reference above, and there were 3 distinct d-band peaks between 0 and 6 eV. No previous Pt reference measurement using MIES has been published, but it is expected the spectrum would look similar to the Ru MIES spectrum above, because the metallic nature of the surface would promote AN+RI de-excitation.

## References

[1] J. Rowe, H. Ibach, Surface and bulk contributions to ultraviolet photoemission spectra of silicon, Phys. Rev. Lett. 32 (1974) 421.
[2] M. Seah, W. Dench, Quantitative electron spectroscopy of surfaces: a standard data base for electron inelastic mean free paths in solids, Surf. Interface Anal. 1 (1979) 2-11.
[3] Y. Harada, S. Masuda, H. Ozaki, Electron spectroscopy using metastable atoms as probes for solid surfaces, Chem. Rev. 97 (1997) 1897-1952.
[4] H. Morgner, The characterization of liquid and solid surfaces with metastable helium atoms, Adv. At. Mol. Opt. Phys. 42 (2000) 387-488.
[5] B.A. Chambers, C.J. Shearer, L. Yu, C.T. Gibson, G.G. Andersson, Measuring the density of states of the inner and outer wall of double-walled carbon nanotubes, Nanomaterials 8 (2018) 448.
[6] G. Krishnan, H.S. Al Qahtani, J. Li, Y. Yin, N. Eom, V.B. Golovko, G.F. Metha, G.G. Andersson, Investigation of Ligand-Stabilized Gold Clusters on Defect-Rich Titania, J. Phys. Chem. C 121 (2017) 28007-28016.
[7] G. Krishnan, N. Eom, R.M. Kirk, V.B. Golovko, G.F. Metha, G.G. Andersson, Investigation of Phosphine Ligand Protected Au13 Clusters on Defect Rich Titania, J. Phys. Chem. C 123 (2019) 6642−6649.
[8] G.G. Andersson, V.B. Golovko, J.F. Alvino, T. Bennett, O. Wrede, S.M. Mejia, H.S. Al Qahtani, R. Adnan, N. Gunby, D.P. Anderson, Phosphine-stabilised Au9 clusters interacting with titania and silica surfaces: The first evidence for the density of states signature of the support-immobilised cluster, J. Chem. Phys. 141 (2014) 014702.
[9] H.D. Hagstrum, Excited-Atom Deexcitation Spectroscopy using Incident Ions, Phys. Rev. Lett. 43 (1979) 1050.
[10] B. Heinz, H. Morgner, A metastable induced electron spectroscopy study of graphite: The k-vector dependence of the ionization probability, Surf. Sci. 405 (1998) 104-111.
[11] H. Haberland, Mall, M., Moseler, M., Qiang, Y., Reiners, T., and Thurner, Y., Filling of micron-sized contact holes with copper by energetic cluster impact, J. Vac. Sci. Technol. A 12 (1994) 2925-2930.
[12] H. Haberland, Karrais, M., Mall, M., and Thurner, Y., Thin films from energetic cluster impact: A feasibility study, J. Vac. Sci. Technol. A 10 (1992) 3266-3271.
[13] V.N. Popok, I. Barke, E.E.B. Campbell, K.-H. Meiwes-Broer, Cluster–surface interaction: From soft landing to implantation, Surf. Sci. Rep. 66 (2011) 347-377.
[14] E.L. Strein, D. Allred, Eliminating carbon contamination on oxidized Si surfaces using a VUV excimer lamp, Thin Solid Films 517 (2008) 1011-1015.
[15] L. Howard-Fabretto, T.J. Gorey, G. Li, S. Tesana, G.F. Metha, S.L. Anderson, G.G. Andersson, The interaction of size-selected Ru3 clusters with RF-deposited TiO2: probing Ru-CO binding sites with CO-Temperature Programmed Desorption, Nanoscale Advances (2021).
[16] D.J. Morgan, Resolving ruthenium: XPS studies of common ruthenium materials, Surf. Interface Anal. 47 (2015) 1072-1079.
[17] Y.J. Kim, Y. Gao, S.A. Chambers, Core-level X-ray photoelectron spectra and X-ray

photoelectron diffraction of RuO2 (110) grown by molecular beam epitaxy on TiO2 (110), Appl. Surf. Sci. 120 (1997) 250-260.
[18] J. Riga, C. Tenret-Noel, J.-J. Pireaux, R. Caudano, J. Verbist, Y. Gobillon, Electronic structure of rutile oxides TiO2, RuO2 and IrO2 studied by X-ray photoelectron spectroscopy, Phys. Scr. 16 (1977) 351.
[19] J. Chastain, Handbook of X-ray photoelectron spectroscopy, Perkin-Elmer Corporation, Minnesota, USA, 1992, pp. 221.
[20] F. Eschen, M. Heyerhoff, H. Morgner, J. Vogt, The concentration-depth profile at the surface of a solution of tetrabutylammonium iodide in formamide, based on angle-resolved photoelectron spectroscopy, J. Phys. Condens. Matter 7 (1995) 1961.
[21] L. Sutton, Tables of interatomic distances and configuration in molecules and ions, Chemical Society1965.
[22] G. Fuentes, E. Elizalde, F. Yubero, J. Sanz, Electron inelastic mean free path for Ti, TiC, TiN and TiO2 as determined by quantitative reflection electron energy-loss spectroscopy, Surf. Interface Anal. 33 (2002) 230-237.
[23] C. Mead, W. Spitzer, Fermi level position at semiconductor surfaces, Phys. Rev. Lett. 10 (1963) 471.
[24] G. Wertheim, S. DiCenzo, Cluster growth and core-electron binding energies in supported metal clusters, Phys. Rev. B 37 (1988) 844.
[25] S. DiCenzo, G. Wertheim, Photoelectron spectroscopy of supported metal clusters, Solid State Phys. 11 (1985) 203.
[26] W. Eberhardt, P. Fayet, D. Cox, Z. Fu, A. Kaldor, R. Sherwood, D. Sondericker, Photoemission from mass-selected monodispersed Pt clusters, Phys. Rev. Lett. 64 (1990) 780.
[27] D.J. Alberas, J. Kiss, Z.-M. Liu, J.M. White, Surface chemistry of hydrazine on Pt (111), Surf. Sci. 278 (1992) 51-61.
[28] J. Crowell, E. Garfunkel, G. Somorjai, The coadsorption of potassium and CO on the Pt (111) crystal surface: A TDS, HREELS and UPS study, Surf. Sci. 121 (1982) 303-320.